\documentclass[journal,onecolumn]{IEEEtran}

\usepackage{datenumber}
\usepackage{amsmath} 
\usepackage{amssymb} 
\usepackage{amsfonts} 
\usepackage{verbatim}  
\usepackage{color} 
\usepackage{booktabs} 
\def\arraystretch{1.6}
\usepackage{nicefrac} 
\usepackage{multirow}
\usepackage{bigstrut}     
\usepackage{array}  
\usepackage{arydshln}  
\usepackage{esvect}  
\usepackage{pgfplots} \usetikzlibrary{plotmarks,patterns}
 \usepackage{url}
 \usepackage{enumerate}
\usepackage{mathtools}
\usepackage{hhline} 
\usepackage{afterpage}
\usepackage{balance}

\newcommand{\Null}{\mathsf{null}}
\newcommand{\Span}[1]{\mathsf{span} \!\left(  #1  \right)}
\newcommand{\RSpan}[1]{\mathsf{rspan} \left( #1 \right)}
\newcommand{\RSpanp}[1]{\mathsf{rspan} \big( #1 \big)}
\newcommand{\Rank}[1]{\mathsf{rank} \big( #1 \big)}
\newcommand{\Dim}[1]{\mathsf{dim} \left( #1 \right)}
\newcommand{\Dimp}[1]{\mathsf{dim} \big( #1 \big)}
\newcommand{\Bdiag}{\mathsf{bdiag}}
\newcommand{\bdiag}[1]{\Bdiag \left( #1 \right)}

\newcommand{\Stack}{\mathsf{stack}}
\newcommand{\stack}[1]{\Stack \left( #1 \right)}
\newcommand{\stackp}[1]{\Stack \Big( #1 \Big)}

\newcommand{\pare}[1] {\left( #1 \right)} 
\newcommand{\parep}[1] {\big( #1 \big)} 
\newcommand{\Min}[1] {\min \parep{#1}} 
\newcommand{\Max}[1] {\max \parep{#1}} 
\newcommand{\sobreq}[2] {\overset{\text{#1}}{#2}} 

\DeclarePairedDelimiter\ceil{\lceil}{\rceil}
\DeclarePairedDelimiter\floor{\lfloor}{\rfloor}

\newcolumntype{C}{>{$}c<{$}} 
\newcolumntype{L}{>{$}l<{$}}
\newcolumntype{R}{>{$}r<{$}}

\newcommand{\0}{\mathbf{0}}
\newcommand{\Ac}{\mathcal{G}}

\newcommand{\AI}{\text{A.I}}
\newcommand{\AII}{\text{A.II}}

\newcommand{\BS}[1]{\Tx{#1}}
\newcommand{\Bmax}{B}
\newcommand{\BI}{\text{B.I}}
\newcommand{\BII}{\text{B.II}}
\newcommand{\CI}{\text{C.I}}
\newcommand{\CII}{\text{C.II}}
\newcommand{\CIII}{\text{C.III}}
\newcommand{\CIV}{\text{C.IV}}
\newcommand{\Rmat}[2]{\!\in\mathbb{R}^{#1 \times #2}}
\newcommand{\Cmat}[2]{\!\in\mathbb{C}^{#1 \times #2}}

\newcommand{\din}{d^{(\text{in})}}
\newcommand{\djin}{d_j^{(\text{in})}}

\newcommand{\djout}{d_j^{(\text{out})}}

\newcommand{\enters}{\mathbb{Z}^+}

\newcommand{\refbox}[2] {\mbox{#1 \ref{#2}}}
\newcommand{\Fn}{\mathbf{F}}
\newcommand{\Fng}[2]{\Fn_{#1}^{\{#2\}}}
\newcommand{\Fc}{\mathcal{F}}
\newcommand{\Fcg}[2]{\Fc_{#1}^{\{#2\}}}

\newcommand{\Gn}{\mathbf{\Omega}}

\newcommand{\Hn}{\mathbf{H}}

\newcommand{\Hns}[2]{\Hn_{#1}^{\left(#2\right)}}

\newcommand{\I}{\mathbf{I}}
\newcommand{\Ic}{\mathcal{I}}

\newcommand{\Rx}[1]{\text{RX}_{#1}}

\newcommand{\Phin}{\boldsymbol{\Phi}}
\newcommand{\roA}{\rho_{\text{A}}}
\newcommand{\roB}{\rho_{\text{B}}}
\newcommand{\rox}{\rho_{\text{x}}}
\newcommand{\roy}{\rho_{\text{y}}}
\newcommand{\roBSR}[1]{\rho_{\text{PSR,#1}}}

\newcommand{\Sigman}{\boldsymbol{\Sigma}}
\newcommand{\Tc}{\mathcal{T}}

\newcommand{\Tx}[1]{\text{TX}_{#1}}
\newcommand{\Tn}{\mathbf{T}}
\newcommand{\UE}[1]{\Rx{#1}}
\newcommand{\un}{\boldsymbol{\theta}}
\newcommand{\Un}{\mathbf{U}}
\newcommand{\Uns}[2]{{\Un}_{#1}^{\left(#2\right)}}

\newcommand{\Vn}{\mathbf{V}}

\newcommand{\Vns}[2]{\Vn_{#1}^{\left(#2\right)}}
\newcommand{\Vnsg}[2]{\bar{\Vn}_{#1}^{\left(#2\right)}}
\newcommand{\wn}{\boldsymbol{\vartheta}}

\newcommand{\xn}{\mathbf{x}}
\newcommand{\Xn}{\mathbf{X}}
\newcommand{\Xc}{\mathcal{X}}
\newcommand{\yn}{\mathbf{y}}
\newcommand{\Yn}{\mathbf{Y}}
\newcommand{\Yc}{\mathcal{Y}}
\newcommand{\Wn}{\mathbf{W}}

\newcommand{\zn}{\mathbf{z}}

\newtheorem{theorem}{Theorem}
\newtheorem{lemma}{Lemma}
\newtheorem{proposition}{Proposition}

\IEEEoverridecommandlockouts

\newcolumntype{C}{>{$}c<{$}} 
\newcolumntype{L}{>{$}l<{$}}
\newcolumntype{R}{>{$}r<{$}}

\definecolor{creme}{RGB}{255, 253, 208}

\newcommand{\linia}[4]{\addplot [color=gray,very thin,solid,forget plot,-stealth] coordinates{ (#1,#2)(#3,#4)};}

\newcommand{\linianf}[4]{\addplot [color=gray,thin,solid,forget plot] coordinates{ (#1,#2)(#3,#4)};}

\newcommand{\poligon}[1]{\fill [color=creme!90, postaction=
    {pattern=north east lines, pattern color = black!15}] #1;}
\newcommand{\textbox}[3]{\node[fill=white, font=\scriptsize] at (axis cs: #1,#2) {#3};} 
\newcommand{\textboxx}[3]{\node[font=\scriptsize ] at (axis cs: #1,#2) {#3};}

\begin{document}

Submitted to the IEEE Transactions on Information Theory.

\copyright 2015 IEEE. Personal use of this material is permitted. Permission from IEEE must be 
obtained for all other uses, in any current or future media, including 
reprinting/republishing this material for advertising or promotional purposes, creating new 
collective works, for resale or redistribution to servers or lists, or reuse of any copyrighted 
component of this work in other works.

\balance
\newpage

\title{DoF-Delay Trade-Off for the $K$-user MIMO Interference Channel With Delayed CSIT} 
%
%
%
\author{Marc~Torrellas,
        Adrian~Agustin,
        and~Josep~Vidal
\thanks{The authors are with the Signal Theory and Communications Department at the Universitat Polit\`ecnica de Catalunya (UPC), Barcelona \{marc.torrellas.socastro, adrian.agustin, josep.vidal\}@upc.edu}
\thanks{This work has been supported by projects TROPIC FP7 ICT-2011-8-318784 (European Commission), MOSAIC TEC2010-19171/TCM, DISNET TEC2013-41315-R (Ministerio de \mbox{Economia} y Competitividad, Spanish Government and ERDFs), and \mbox{2014SGR-60} (Catalan Administration).}
\thanks{Part of this material was presented at the International Conference on Acoustics, Speech and Signal Processing, Florence, Italy, May 2014 along the papers \cite{TAV} and \cite{TAV_RIA}.}}
 
%
%

\markboth{}%
{Torrellas \MakeLowercase{\textit{et al.}}: DoF-Delay Trade-Off for the $K$-user MIMO Interference Channel With Delayed CSIT}


\maketitle

\begin{abstract}
The degrees of freedom (DoF) of the $K$-user multiple-input multiple-output (MIMO) interference channel are studied when perfect, but delayed channel state information is available at the transmitter side (delayed CSIT). Recent works have proposed schemes that achieve increasing DoF values, but at the cost of long communication delays. This work proposes three linear precoding strategies, formulated in such a way that the achievable DoF can be derived as a function of the transmission delay, thus elucidating its achievable DoF-delay trade-off. All strategies are based on the concept of interference alignment, and built upon three main ingredients: {\it delayed CSIT precoding, user scheduling}, and {\it redundancy transmission}. In this respect, the interference alignment is realized by exploiting delayed CSIT in order to align the interference at the non-intended receivers along the space-time domain. 
Finally, the latter part of this work settles that all the proposed strategies work also for constant channels, except for SISO. In such a case, the schemes can be made feasible by resorting to asymmetric complex signaling concepts. This conclusion removes the time-varying channels assumption unnecessarily common along all the literature on delayed CSIT. \end{abstract}

\begin{IEEEkeywords}
Delayed Channel State Information, Interference Channel, MIMO, Degrees of freedom, Interference Alignment
\end{IEEEkeywords}

\section{Introduction}
%
%

\IEEEPARstart{C}{haracterization} of the degrees of freedom (DoF, also known as the \textit{multiplexing gain}) for interference networks have attracted many researchers during the last decade. They represent the scaling of channel capacity with respect to (w.r.t.) the signal-to-noise ratio (SNR) at the high SNR regime. Since in general capacity expressions are not known, characterizing the DoF sheds some light about how e.g available channel state information (CSI), number of transmit or receive antennas, impact on system capacity. In this context, \textit{interference alignment} (IA) emerged a few years ago as a new tool for managing the signal dimensions (time, frequency, space) in pursuit of attaining the optimal DoF \cite{Maddah-Ali2008}\cite{CJ}. The concept consists on designing the transmitted signals in such a way that they are overlapped (or \textit{aligned}) at the non-intended receivers. Therefore, the interference lies on a reduced dimensional subspace, releasing some dimensions to allocate desired signals which can be retrieved by means of zero-forcing (ZF) concepts.

The idea of IA arose in the context of index coding in \cite{Birk&Kol},
while its application to wireless networks crystallized about ten years later for the 2-user multiple-input multiple-output (MIMO) X-channel in \cite{Maddah-Ali2008} and for the $K$-user single-input single-output (SISO) interference channel (IC) with $K>2$ in \cite{CJ}. This latter reference entailed a breakthrough since the authors proposed a linear scheme providing \textit{each user half the cake} as compared to the single-user case, i.e. a total of $\frac{Km}{2}$ DoF are achieved over the network when each node is equipped with $m$ antennas. However, for SISO ($m=1$) their scheme requires $2^{K^2}$ time slots and applies only for time-varying channels. This latter issue has been partially solved by means of asymmetric complex signaling (ACS) concepts \cite{ACS}, i.e. exploiting improper Gaussian signaling. 
In the literature the benefits of ACS have been shown in terms of : 1) boosting the DoF of the IC with constant channel \cite{ACS}\cite{TAV_ppM1}\cite{ACS_4users}; 2) improving the sum-rate of the system \cite{ACS_sumrate}\cite{SandraICASSP}\cite{SandraGlobecom}; and 3) reducing the total transmitted power for a given QoS \cite{ACS_reducedPower}. 

Since its emergence, IA has become a very useful tool for studying many multi-user scenarios in combination with the well-known null-steering or ZF approach \cite{ZFBD}. A very extensive survey of IA applications can be found in \cite{IAtutorial}. Of particular interest for this work is the characterization of the DoF of the MIMO IC for three users \cite{AlignmentChainsTrans}\cite{TAV_ppM1} and more than 3 users \cite{Ghasemi2010}\cite{GenieChainsArxiv}.
Moreover, there exists another type of IA in the literature, denoted as ergodic IA \cite{ergodicIA}, that exploits opportunistically channel variations to perform IA. Basically, it consists in using the channel once and then wait for a particular channel realization satisfying some conditions that allow canceling the interference. Nevertheless, although it performs better than conventional IA at low-medium SNR regime, this approach becomes more a theoretical than a practical result, since the average delay expresses approximately as $\gamma^{K^2}$, with $\gamma$ denoting the SNR.

All IA-based and ZF-based schemes previously mentioned require perfect and instantaneous channel state information at the transmitters (CSIT), an assumption not always realistic in wireless cellular networks. For example, in frequency division duplexing systems, channel coefficients are usually estimated at the receivers by means of a training period, and then fed back to the transmitters, which incurs in delays and errors. The feedback error has been widely studied in the literature, and the main conclusion is that in order to preserve the DoF, the number of quantization bits should scale with the logarithm of the SNR \cite{Jindal}.  
On the other hand, it is usually assumed a block fading channel model, where channel remains constant in blocks of duration equal to the channel coherence time. Consequently, when the feedback delay is higher than the coherence time, the available CSIT is completely outdated. In such a case, all strategies based on full CSIT are no longer effective to handle the interference. 
 
{In this respect, Maddah-Ali and Tse (MAT) introduced in \cite{MAT} a new framework where IA concepts can be exploited even when the CSIT is completely outdated, referred to as delayed CSIT. Indeed, they assume {\it perfect} delayed CSIT, which is more realistic since during the time elapsed between transmissions, receivers can report a sufficient number of quantization bits. However, there are some drawbacks, e.g. {since the current channels are not known, the effective rate at which symbols are sent is simply based on statistics and the particular topology/setting.} Moreover, notice that the concept of reciprocity \cite{reciprocity}\footnote{{This property states that if the role of transmitters and receivers are switched, the DoF are preserved}. Obviously, this property applies only when exactly the same CSI is available at both sides, e.g. with full CSIT.} no longer holds for networks restricted to only delayed CSIT, which in case of full CSIT severely simplifies the challenge of DoF characterization for all MIMO settings.}
 
The MAT scheme was the first application of IA concepts using only delayed CSIT. Originally proposed for the $K$-user MISO broadcast channel (BC), the communication is carried out along $K$ phases for transmitting $b$ symbols per user. The two main ingredients of their approach are {\it delayed CSIT precoding} and {\it user scheduling}. On the one hand, linear combinations of all $b$ symbols exploiting the delayed CSIT are sent along all the phases, working similarly to the automatic repeat request (ARQ) protocols, where the same message (or packet) is retransmitted until it can reliably be decoded at the receiver side. On the other hand, the scheme imposes that during each phase $p$ users are served in different time instances by groups of $p \leq K $ users, whereas the rest of users listen and learn about the interference. This user scheduling is decided beforehand by protocol and independently of channel realization, and the objective is to control the number of interference terms contributing to the signal observed at each receiver.
For example, during the first phase, users are served in a TDMA fashion $(p=1)$, i.e. first the transmitter sends the symbols of user one, then symbols of user two, and so on. 
The scheme is designed for a MISO setting, with the number of transmitted symbols higher than the receive antennas, thus symbols cannot be linearly decoded after the first phase. However, under the assumption that channels are uncorrelated across users, all users (served and listening) obtain different and independent linear combinations (LCs) of each set of symbols after the first phase. The obtained LCs when one user acts as listening user (thus containing non-intended symbols) will be denoted as {\it overheard interference}. They are known at one receiver at least, and desired by another one. Then, the objective of the following phases is to exploit the delayed CSIT to reconstruct this overheard interference, and then retransmit it since it will be removable for the receivers that already know it. This idea allows that more than one user can be simultaneously served after the first phase.

Inspired by the MAT scheme in \cite{MAT}, some works appeared for studying the interference channel with delayed CSIT. However, while the 2-user MIMO IC was completely characterized by Vaze et al. in \cite{Vaze2IC}, having $K>2$ users is still an open problem. Basically, this is because in the MIMO IC, in contrast to the MIMO BC, each transmitter has only access to its own symbols, thus it can only reconstruct part of the overheard interference. Existing contributions for the $K$-user $(M,N)$ MIMO IC with delayed CSIT are next reviewed, i.e. with $M$, $N$ antennas at the transmitters, receivers, respectively.

For the case $M \geq K$, Torrellas et al. proposed in \cite{TAV} a linear precoding scheme achieving $\frac{2}{K+1}$ DoF per user\footnote{This result was independently derived in the PhD Thesis \cite{PhD_ghasemi}.}. This is accomplished by delivering $K$ symbols to each user after a two-phase transmission protocol of duration $K + \binom{K}{2}$ time slots. The first phase is developed in a TDMA fashion $(p=1)$,
whereas in the second phase only one pair of transmitters is simultaneously active $(p=2)$. Then, it can be interpreted as the application of the MAT scheme tools ({\it delayed CSIT precoding} and {\it user scheduling}), to the IC, constrained to the use of only two phases. This is because aligning more than 2 users at one receiver using only delayed CSIT is not straightforward when transmitters are distributted and have only delayed CSIT.

A different approach was proposed by Maleki et al. in \cite{MalekiRIA} for the 3-user SISO IC. Their two-phase scheme, denoted as the retrospective interference alignment (RIA) scheme, provides 3 symbols to each user after a transmission protocol of duration 8 time slots, thus attaining $\frac{3}{8}$ DoF per user. In contrast to the MAT scheme, no scheduling is applied, and all transmitters are active and interfering each other during all the communication time $(p=K)$. {The main innovation of \cite{MalekiRIA} lies on performing a first phase transmission with \textit{redundancy}, i.e. the receiver obtains more linear combinations than the number of symbols of one user. This allows processing the signal at the receiver side to project it onto singular vector spaces where the desired signals are only interfered by the symbols of one user, i.e. the contribution of one set of non-intended symbols can be removed.
Thanks to this {\it partial interference nulling}, interference is easily aligned during the second phase by exploiting delayed CSIT. It is worth pointing out that in contrast to the MAT scheme where desired signals and overheard interference are acquired from different time instants, now they are obtained together, thus after the first receivers have no interference-free LCs of desired signals.
Two extensions followed \cite{MalekiRIA}: \cite{RIA_Kusers} and \cite{TAV_RIA}. First, Maggi et al. proposed in \cite{RIA_Kusers} a generalization of the concept for $K>3$ users, even though their main conclusion was that it is preferable to consider only 3 active transmitter-receiver pairs and apply time-sharing. And second, Torrellas et al. propose in \cite{TAV_RIA} generalization for the 3-user MIMO case, improving the state-of-the-art for certain antenna settings.

Combining all the ingredients in \cite{MAT} and \cite{MalekiRIA} ({\it delayed CSIT precoding, user scheduling}, and {\it redundancy transmission}), Abdoli et al. proposed in \cite{Abdoli_IC} a precoding scheme for the $K$-user SISO IC. This scheme will be referred hereafter as the precoding, scheduling, redundancy (PSR) scheme. Developed in $K$ phases, its sum DoF increase with the number of users $K$, thus differing from the rule of thumb provided in \cite{RIA_Kusers}. Moreover, the authors conjectured that in contrast to the full CSIT case, the sum DoF of the IC with delayed CSIT collapse to a constant value as the number of users becomes asymptotically high. More recently, a generalization of the PSR scheme has been proposed by Hao et al. in \cite{Hao_BSR_MISOIC} for the $K$-user MISO IC, with transmitters equipped with $K-1$ or more antennas and single-antenna receivers.

The inner bounds proposed in \cite{Abdoli_IC} and \cite{Hao_BSR_MISOIC} outperform any proposed work for each setting, although at the cost of a long communication delay with limited DoF gains. For example, \cite{MalekiRIA} obtains $\frac{3}{8}$ DoF per user in the 3-user SISO IC with only 8 slots, whereas the scheme in \cite{Abdoli_IC} requires 31 slots to increase the achieved DoF to $\frac{12}{31}$, i.e. a 3\% of DoF gain. Moreover, it requires 3 phases, thus more uplink resources dedicated for channel feedback. Similarly, for the 6-user MISO IC with 6 antennas at the transmitters, the scheme in \cite{Hao_BSR_MISOIC} provides a 10\% DoF gain w.r.t. \cite{TAV}, but this is at the cost of 1422 instead of 21 slots. Summarizing, it seems that the best schemes in terms of DoF require a long transmission time, providing no significant gains w.r.t. other schemes with shorter transmission time. {In this respect, the DoF-delay trade-off comes up as an interesting topic to be investigated, i.e. comparison of schemes not only takes into account the achievable DoF, but also the transmission duration.}

{Ergodic IA concepts have been also extended to the case of delayed CSIT in \cite{ErgodicIA_dCSIT}. However, although the DoF results provided in \cite{ErgodicIA_dCSIT} outperform some of the material presented here, it has some drawbacks. First, it is assumed that transmitters wait until channels satisfy some conditions and then transmit. This entails very long delays, which is not desirable in practical terms as explained above. And second, ergodic IA (also when applied to the delayed CSIT setting) relies on channel variations, thus it is only suitable for time-varying channels. This latter assumption is common throughout all the literature on delayed CSIT, and it follows from 
understanding that channel feedback will incur a delay larger than channel coherence time. But, is this assumption necessary indeed? Note that in practice the transmitter has no way to know the current channel coefficients. Therefore, one may ask which of the state-of-the-art results are applicable in case there is delayed CSIT, the channel remains constant, and transmitter is not aware of this, thus performing a delayed CSIT strategy anyways. 

Finally, it is worth pointing out that IA concepts have also been extended to the case of no CSIT. This kind of IA is labeled in the literature as Blind IA \cite{BlindIA}. In such a case, proper channel variations are chosen for interference alignment. Although it was initially assumed that they appear naturally \cite{BlindIAprimer}, i.e. by proper user selection, some recent works have shown that they can be manipulated or artificially constructed from a constant channel by means of reconfigurable antennas \cite{BlindIA}. In this work, Blind IA will not be considered since contrary to Ergodic IA it requires constant channels and reconfigurable antennas, whereas the schemes proposed here can be used either for constant or time-varying channels.

\subsection{Contributions}
\label{sec:contributions}

This work studies the $K$-user $(M,N)$ MIMO IC with delayed CSIT, see \mbox{Fig. \ref{fig:scenario}}, where transmitters and receivers are equipped with $M$, $N$ antennas, respectively. This paper subsumes our two previous conference papers \cite{TAV}\cite{TAV_RIA} on this subject, extending the proposed schemes to the general $K$-user MIMO case. These and the rest of contributions of this work are summarized next:

\begin{figure}[]
\begin{minipage}[]{1\linewidth}
  \centering 
  \centerline{\includegraphics[width=0.45\linewidth]{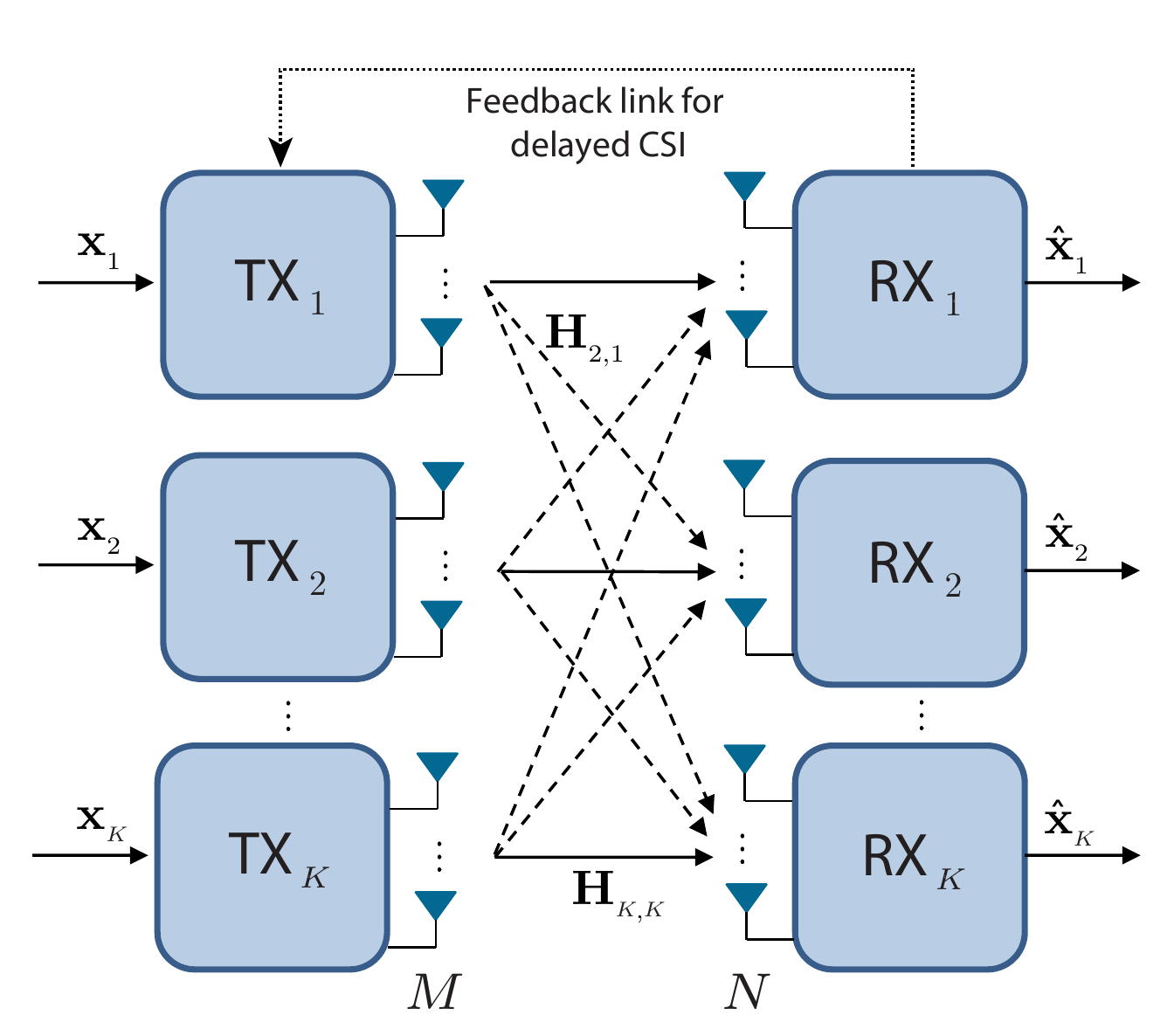}}        
\end{minipage}  
\caption{The $K$-user MIMO IC, with $(M,N)$ antennas at the transmitters and receivers, respectively. Solid lines define the intended signals, while dotted lines denote the interfering signals.}
\label{fig:scenario}
\end{figure}

\begin{itemize}
  \item When $M\leq N$, new DoF inner bounds are provided by generalization of the RIA scheme in \cite{MalekiRIA} to the $K$-user MIMO case. 
  In contrast to the rule of thumb in \cite{RIA_Kusers}, it is shown that considering $L \in \{3,4,\dots,K\}$ users simultaneously active may increase the attained DoF, where the optimal value of $L$ depends on each antenna setting and the total number of users $K$. 
  
  \item When $M>N$, new DoF inner bounds are provided by generalization of the two-phase scheme in \cite{TAV} to the $K$-user MIMO case. While in the original scheme the second phase was developed by rounds with only two active users, here groups of $G_2 \in \{2,\dots,K\}$ users are active during the second phase. Then, $G_2-1$ interference terms are generated {per} round at each active receiver. This imposes a number of IA constraints that are feasible depending on the antenna setting. In this respect, the optimal value of $G_2$ is designed according to the antenna setting and the number of users $K$ in order to obtain the maximum DoF. Inspired by the way it is carried out, we denote this scheme as the TDMA groups (TG) scheme.
   
  \item When $M\approx N$ and $K=3$ users, new DoF inner bounds are obtained by generalization of the PSR scheme in \cite{Abdoli_IC} to the MIMO case. Moreover, this scheme also improves previous inner bounds when it is applied in a $K$-user MIMO IC combined with time-sharing concepts.
  
 \item The DoF-delay trade-off of proposed and state-of-the-art schemes is studied. One of the main conclusions is that for most cases the number of transmitted symbols and duration of the phases can be severely reduced for the sake of reducing communication delay without significant DoF losses. 
  Those parameters are derived for each scheme by formulating a DoF maximization problem, which allows obtaining DoF inner bounds as a function of each setting, i.e. number of users and antenna configuration. Then, our DoF-delay analysis is carried out by including additional constraints in pursuit of bounding these parameters. {As an example, for $(M,N,K)=(4,1,6)$ the TG scheme requires 75 slots, whereas a modified version of this scheme can be derived losing a 3\% of the achievable DoF w.r.t. the unbounded case but requiring 27 slots only.}
 
 \item {The assumption that channels should be uncorrelated in time when using delayed CSIT schemes is settled as unnecessary for all cases except for SISO. In such a case, we prove that: 1) the schemes in the literature fail, although 2) they can be made feasible by resorting to asymmetric complex signaling concepts}.
 
\end{itemize}


\subsection{Organization}
The paper is organized as follows. \mbox{Section \ref{sec:systModel}} introduces the system model considered in this work. Next, Section \ref{sec:MainResults} summarizes the main results: DoF inner bounds for the $K$-user MIMO IC with delayed CSIT with time-varying or constant channels. DoF inner bounds are attained by means of three precoding schemes: RIA, TG, and 3-user PSR. Basically, the difference between the first two approaches is how the overheard interference is obtained. In RIA, described in \mbox{Section \ref{sec:innerboundRIA}}, all users are active and the overheard interference is acquired by exploiting {\it redundancy transmission}. In contrast, the TG scheme described in \mbox{Section \ref{sec:innerboundTG}} is more similar to the MAT scheme, and the first phase is carried out orthogonally, thus the overheard interference is individually observed. Finally, the PSR scheme described in \mbox{Section \ref{sec:PSR}} generates the first phase overheard interference as the RIA scheme, but consists of three phases including all the ingredients: {\it delayed CSIT precoding, user scheduling}, and {\it redundancy transmission}. The MIMO generalization of those schemes is obtained through the formulation of a DoF maximization problem, providing the best system parameters for each scheme given the number of users and antenna setting. Also, this formulation allows studying the DoF-delay trade-off of the proposed schemes in \mbox{Section \ref{sec:practical}}. Next, \mbox{Section \ref{sec:constant}} addresses the analysis of delayed CSIT schemes under constant channels. Finally, conclusions and future work are drawn in \mbox{Section \ref{sec:conclusion}}.

\subsection{Notation}

Boldface and lowercase types denote column vectors ($\mathbf{x}$). Boldface and uppercase types are used for matrices ($\mathbf{X}$). Sets and subspaces are denoted by uppercase types written in calligraphic fonts ($\mathcal{X}$). Furthermore, $\mathbb{C}$ and $\enters$ denote the field of complex numbers, and positive integers, respectively.

We define $\0$ and $\mathbf{I}$ as the all-zero and identity matrices, respectively, with suitable dimensions according to the context. For vectors and matrices, $(\cdot)^T$ is the transpose operator, and $(\cdot)^H$ is the transpose and conjugate operator. Moreover, the following two predefined vector and matrix operations are defined:
\begin{IEEEeqnarray*}{c c c}
\def\arraystretch{1.2}
\stack{\Xn,\Yn} =
\begin{bmatrix}
\Xn \\
\Yn \\
\end{bmatrix},
& \quad &
\bdiag{\Xn,\Yn} =
\begin{bmatrix}
	\Xn & \0 \\
	\0 & \Yn \\
\end{bmatrix}.
\end{IEEEeqnarray*}

$\Span{\Xn}$ is usually used to define the \textit{column subspace}, containing all possible linear combinations of the columns. However, in this work we always use the \textit{row subspace}, defined as ${\Xc = \RSpan{\Xn}=
\mathsf{span} \big( \Xn^T \big) }$, whose dimension is given by 
$\Dim{\Xc}={\Rank{\Xn}}$. 
In this regard, three operations between subspaces (or in general for sets) are defined: $\Xc \cap \Yc$ defines the \textit{intersection subspace}, given by the elements that belong to both $\Xc$ and $\Yc$; $\Xc + \Yc$ defines the \textit{sum subspace}, containing all the elements that can be generated linearly combining the elements of $\Xc$ and $\Yc$; and finally, $\Xc \backslash \Yc$ contains the elements that belong to $\Xc$ but not to $\Yc$. 

Notice that the operator $\Stack{(\cdot)}$ produces a matrix whose rows lie on the sum of row subspaces, i.e. lying on $\Xc + \Yc$, whereas a basis for the intersection subspace can be bound by exploiting the fact that operations over Linear Subspaces form a Boolean Algebra. In this regard, let $\bar{\Xc}$ denote the subspace complementary to $\Xc$, and consider the $N$ subspaces $\Xc_1,\dots,\Xc_N$, then the following holds:
\begin{IEEEeqnarray}{c}
 \bigcap_{i=1\dots N} \!\!\! \Xc_i = \overline{\sum_{i=1\dots N} \!\!\!\!\bar{\Xc}_i}. 
 \label{eq:intersection}
\end{IEEEeqnarray}

\section{System model}
\label{sec:systModel}

The $K$-user MIMO IC consists of $K$ transmitter-receiver pairs {sharing the same frequency band} and coexisting in the same area, see Fig. \ref{fig:scenario}. Communication is carried out in $P\leq3$ phases, with each phase $p$ in turn divided in $R_p$ rounds of duration $S_p$ time slots each, see \mbox{Fig. \ref{fig:frame}}. The total number of slots used for data transmission are
\begin{IEEEeqnarray}{c}
\tau= \sum_{p=1}^P \tau_p \,, \quad \tau_p = R_p S_p.
\end{IEEEeqnarray}
Each transmitter ($\Tx{i}$) is equipped with $M$ antennas, and wants to deliver $b$ independent symbols to receiver ($\Rx{i}$), equipped with $N$ antennas. One of the key parameters defining this channel is its {\it antenna ratio}, defined as follows:
\begin{IEEEeqnarray}{c}
 \rho = \frac{M}{N},
\end{IEEEeqnarray}
Note that $P$, $b$, $R_p$, and $S_p$ are designed as a function of $\rho$ and $K$, and will be detailed later for each precoding strategy.

\subsection{Signal Model}

During the $(p,r)$th round, i.e. round $r$ of phase $p$, only a specific group of users denoted by $\Ac^{(p,r)}$, is served. All groups of each phase have the same cardinality, with $G_p = \left| \Ac^{(p,r)} \right| , \forall r$. According to these definitions, the signal received at $\Rx{j}$ writes as
\begin{IEEEeqnarray}{c}
\yn_j^{\left(p,r\right)} = \!\!\! \sum_{i \in \Ac^{(p,r)}} \!\!\!\!\Hns{j,i}{p,r}
 \Vns{i}{p,r} \mathbf{x}_i + \mathbf{n}_j^{(p,r)},
\label{eq:SystemModel}
\end{IEEEeqnarray}
where $\yn_j^{\left(p,r\right)} \Cmat{NS_p}{1}$ is the vector containing the signals observed at the $j$th receiver, $\mathbf{x}_i \Cmat{b}{1}$ contains the $b$ uncorrelated unit-powered complex-valued data symbols intended to the $i$th receiver. Note that linear combinations of the same $b$ symbols are transmitted during all phases, but receivers will not be able to decode them until the end of the communication either because the reduced number of receive antennas, or because of interference. Besides, $\Vn_i^{\left(p,r\right)}\Cmat{MS_p}{b}$ denotes the precoding matrix carrying the signals desired by the $i$th user, designed subject to a maximum transmission power per user $\gamma$, and with ${\Vn_i^{\left(p,r\right)}=\0,\forall i \notin \Ac^{(p,r)}}$, and $\mathbf{n}_j^{(p,r)} \Cmat{NS_p}{1}$ is the unit-powered noise term. Since the focus of this paper is on DoF analysis, all noise terms will be omitted, thus the maximum transmission power $\gamma$ denotes also the SNR.

\begin{figure}[]
\begin{minipage}[]{1\linewidth}
  \centering  
  \centerline{\includegraphics[width=0.45\linewidth]{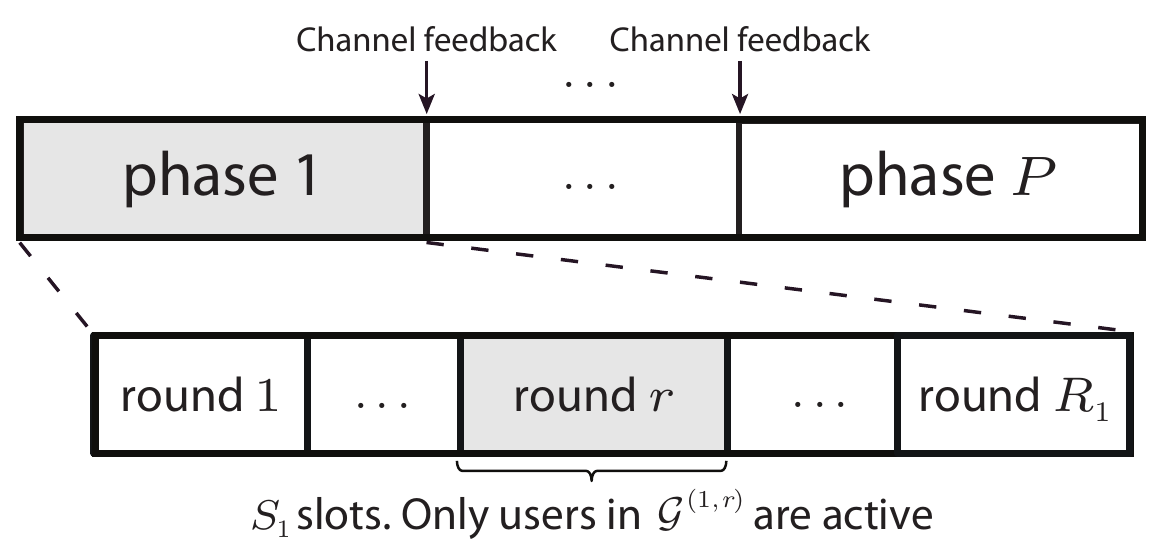}}
\end{minipage}    
\caption{General protocol frame of the proposed schemes. After each phase, CSI is obtained at the transmitters through feedback (FB). There are $P\leq3$ phases, where the phase $p$ is divided in $R_p$ rounds. During each round $r$ one different group of $G_p$ transmitters is served, predefined by the set $\Ac^{(p,r)}$. Moreover, each round of the phase $p$ is in turn divided in $S_p$ time slots.}
\label{fig:frame}  
\end{figure} 

The channel coefficients for each slot and each link between transmitter and receiver are described by an $N\times M$ matrix. Then, the channel matrix $\Hn_{j,i}^{\left(p,r\right)} \Cmat{NS_p}{MS_p}$ in (\ref{eq:SystemModel}) is formed as the block diagonal composition of $S_p$ of such matrices, thus contains the channel gains from antennas of $\Tx{i}$ to $\Rx{j}$ during all time slots of the $(p,r)$th round. 

{Usually, most works on delayed CSIT assume a flat block fading channel model, i.e. channels are i.i.d. as $\mathcal{CN}\!\pare{0,1}$, and completely uncorrelated in time and space. This will be the setting for all this work, except for \mbox{Section \ref{sec:constant}}, where the objective is to show that the proposed precoding schemes work without assuming time-varying channels.}

After each phase $\Rx{j}$ collects all the received signals and process them by means of the linear filter $\Un_j^{(p)} \Cmat{\beta_p}{N\tau_p}$, where $\beta_p$ and the design of those filters will be detailed for each case. The processed signal vector for phase $p$ writes as
\begin{IEEEeqnarray*}{c}
 \zn_j^{(p)}= \Un_j^{(p)} \, \stack{\yn_j^{(p,1)},\dots,\yn_j^{(p,R_p)}}.
\end{IEEEeqnarray*}
 Similarly, with the objective of retrieving $b$ linear combinations of its desired symbols, each receiver collects the signals along all the communication. Therefore, by grouping the magnitudes of the different rounds and phases the global-input output relationship is written in compact form as
\begin{IEEEeqnarray}{c}
\label{eq:SystemModelextended}
\def\arraystretch{1.2}
\begin{matrix}
\zn_j = \stack{\zn_j^{(1)},\dots,\zn_j^{(P)}} =  \Gn_j \left[\xn_1^T,\dots,\xn_K^T \right]^T, \\[2mm]
	\Gn_j = \Un_j
    \begin{bmatrix} 
		\Hn_{j,1} \Vn_{1}, \, \ldots ,\,
		\Hn_{j,K} \Vn_{K}
	\end{bmatrix}, \\[2mm]
	\Un_j= \bdiag{\Un_j^{(1)}, \dots, \Un_j^{(P)}} \\[2mm]
\Hn_{j,i} = \bdiag{
\begin{matrix}
	 \Hn_{j,i}^{\left(1\right)}, \dots,
	\Hn_{j,i}^{\left(P\right)}
\end{matrix} } , \quad
\Vn_{i} = \stack{
\begin{matrix}
	\Vn_{i}^{\left(1\right)}, \dots,
	\Vn_{i}^{\left(P\right)}
\end{matrix} }, \\[2mm]
\Hn_{j,i}^{(p)} = \bdiag{
\begin{matrix}
	\Hn_{j,i}^{\left(p,1\right)}, \,
	\Hn_{j,i}^{\left(p,2\right)}, \,\ldots, \,
	\Hn_{j,i}^{\left(p,R_p\right)}
\end{matrix} }, \\[2mm]
\Vn_{i}^{(p)} = \stack{
\begin{matrix}
	\Vn_{i}^{\left(p,1\right)}, \,
	\Vn_{i}^{\left(p,2\right)}, \,\ldots, \,
	\Vn_{i}^{\left(p,R_p\right)}
\end{matrix} },
\end{matrix}
\end{IEEEeqnarray}
\\[-2mm]
where $\Gn_j$ is the signal space matrix \cite{TAV_ppM1}, defining the subspaces occupied by the received signals at each receiver, $\Un_{j}$ is the composition of all per-phase receiving filters $\Un_j^{(p)}$ whose dimensions depend on each precoding scheme,
${\Hn_{j,i} \Cmat{N\tau}{M\tau}}$, ${\Vn_i\Cmat{M\tau}{b}}$, ${\Hn_{j,i}^{(p)} \! \Cmat{N\tau_p}{M\tau_p}}$, and ${\Vn_{i}^{(p)} \Cmat{M\tau_p}{b}}$. 

All precoding and receiving filters are designed subject to a delayed CSIT model. Using the given formulation, this means that only the channels
\begin{IEEEeqnarray*}{c}
 \{ \Hns{j,i}{\varrho}\}_{\varrho=1}^{p-1},\forall i,j,
\end{IEEEeqnarray*}
are available at the transmitter side at the beginning of the phase $p$, whereas all CSI is instantaneously assumed to be known at the receiver side.

\subsection{Degrees of freedom}

We analyze the normalized DoF per user, i.e. divided by the number of receive antennas, given by \cite{Cover}
\begin{IEEEeqnarray}{c}
\djin \! = \lim_{\gamma \to \infty} \,\, \frac{C_{\Sigma}(\gamma)}{KN\tau \log_2 \gamma} \leq \djout,
\label{eq:DoFdef}
\end{IEEEeqnarray} 
where $C_{\Sigma}(\gamma)$ denotes the sum capacity for SNR equal to $\gamma$, and $\djout$ denotes the normalized DoF per user outer bound. For $\rho < \frac{1}{K-1}$, the DoF with full CSIT can be achieved without CSIT by applying zero-forcing concepts at the receiver only, see for example Section V.A of \cite{AlignmentChainsTrans}. For the rest of cases, and for comparison purposes, we use the following DoF outer bound:
\\

\begin{minipage}{\linewidth}
\begin{theorem}[DoF Outer bound \cite{GenieChainsArxiv,MAT}]
For the $K$-user MIMO IC with delayed CSIT and antenna ratio $\rho$, the normalized DoF per user are bounded above by:
\\[-10mm]
 \label{th:outerbound} 
\end{theorem}
\begin{center}
\begin{IEEEeqnarray}{c}
\djout = \left\{ \,
\begin{matrix}
      \dfrac{K-1}{K}\rho & \quad & 
      \dfrac{1}{K-1} \leq  \rho  < \alpha \\[4mm]
      \dfrac{\rho}{\rho+1}  & \quad &	 
      \alpha \leq  \rho < \frac{1}{\beta}	    \\[4mm]
      \dfrac{1}{\beta+1} & \quad &  
      \rho  \geq \frac{1}{\beta}
\end{matrix}
\right.
\label{eq:example_left_right1}
\end{IEEEeqnarray}
\\[3mm]
%
%
%

    
\end{center}
\end{minipage}
where $\alpha = \dfrac{K-2}{K^2-3K+1} $ and $\beta = \dfrac{1}{2} + \dfrac{1}{3} + \dots +\dfrac{1}{K}$. \\
\begin{IEEEproof}  
 The first two bounds follow by assuming full CSIT and applying the results in \cite{GenieChainsArxiv}, since this cannot decrease the capacity of a network with delayed CSIT. Similarly, the other bound is based on the idea that cooperation can never hurt the DoF, thus the bounds for the 3-user BC with delayed CSIT in \cite{MAT} can be applied here. \\
\end{IEEEproof}   
\vskip 3mm


Fortunately, the achievable DoF can be written in a more handy way by using standard derivations \cite{Cover}. Consider a receiving filter ${\Wn_j \Cmat{b}{\tau}}$ such that 
\begin{IEEEeqnarray}{c}
\mathbf{{W}}_{j} \Un_j \mathbf{{H}}_{j,i} \mathbf{{V}}_i  = \0 , \quad \forall i \neq j ,
\label{eq:noInterferenceb}
\end{IEEEeqnarray} 
i.e. acting as a linear zero-forcing filter that projects the received signals onto the orthogonal-to-interference space, thus separating desired signals from interference. Then, defining the equivalent channel for $\Rx{j}$ as
\begin{IEEEeqnarray}{c}
 \Hn_j^{(\text{eq})} = \Wn_j \Un_j \Hn_{j,j} \Vn_j, \label{eq:HeqDef}
\end{IEEEeqnarray}
the normalized achievable DoF express as
\begin{IEEEeqnarray}{c}
\label{eq:achDoF}
\djin = 
\frac{1}{N\tau} \Rank{ \Hn_j^{(\text{eq})}}  
\overset{(a)}{\leq}
\frac{b}{N\tau} \leq \djout,
\end{IEEEeqnarray}
where inequality $a$ is satisfied with equality only if after projection the equivalent channel has rank $b$. In other words, after projection each receiver should be able to retrieve $b$ independent and free of interference LCs or observations of its desired symbols. Since usually the precoding matrices are designed to manage the interference, direct channels do not take part on the precoding matrix design. Therefore, it is conjectured that since channels are generic inequality $a$ will be satisfied with equality with probability one. However, for some cases this is not always true, and a rigorous proof is required, as in \cite{TAV_ppM1}.

\subsection{Time-sharing}
\label{sec:timesharing}

Any scheme working for $L<K$ users can be used for the $K$-user MIMO IC by subsequently selecting only $L$ users and turning off the $K-L$ additional users, such that all possibles groups of $L$ users are served once. Let assume that one scheme provides $\tilde{d}_j^{(\text{in})}$ DoF to each of $L$ users along $\tilde{\tau}$ slots. Then, the equivalent DoF per user and duration of the communication when it is used for the $K$-user case write as
\begin{IEEEeqnarray}{c}
 \djin = \frac{L}{K}\tilde{d}_j^{(\text{in})},\quad 
 \tau = \binom{K}{L} \tilde{\tau}.
 \label{eq:timesharing}
\end{IEEEeqnarray}

\section{Main results}
\label{sec:MainResults}

The main results of this work are Theorem \ref{th:innerbound3}, Theorem \ref{th:innerbound}, providing DoF inner boudns for the 3-user and $K$-user MIMO IC, respectively, and Theorem \ref{th:constant}, extending the results to the constant channel case. The first two are next stated and illustrated by means of some examples:

   \pgfplotsset{
    plotoptsPrevIn/.style={color=blue, very thick, dotted},
    plotoptsPropIn/.style={color=orange, thick,},
    plotoptsUp/.style={color=red, thick,dashed},
    plotoptsNoCSIT/.style={color=black, thick, dash pattern=on 3pt off 1pt on 1pt off 1.5pt},
    }

\vskip 5mm

\begin{minipage}{\linewidth}
\begin{theorem}[DoF Inner bound for 3 users]
For the $3$-user MIMO IC with delayed CSIT and antenna ratio $\rho$, the following DoF per user can be achieved:
 \label{th:innerbound3} 
\end{theorem}
\def\arraystretch{1.6}
\begin{IEEEeqnarray*}{c}
\djin = \left\{ \,
\begin{matrix}
      \dfrac{\rho^3}{2-\rho} & \quad & 
      \frac{1}{2} < \rho \leq \roBSR{1} \\[4mm]
      \dfrac{2\rho^2}{5\rho^2-10\rho+8} & \quad & 
      \roBSR{1} < \rho \leq \roBSR{2} \\[4mm]
      \dfrac{6\rho}{3\rho+10} & \quad &  
      \roBSR{2} < \rho < \frac{4}{5} \\[4mm]
      \dfrac{12}{31} & \quad &  
      \rho \geq \frac45
\end{matrix}
\right.
\end{IEEEeqnarray*}
\end{minipage} 
\vskip 5mm
where
\begin{IEEEeqnarray}{c}
 \roBSR{1} = \frac{1}{15}\pare{10 + 5^{2/3}
 \pare{\sqrt[3]{2\pare{3\sqrt{6}+2}} - \sqrt[3]{2\pare{3\sqrt{6}-2}}}
 } \approx 0.7545 \label{eq:roBSR1} \\
 \roBSR{2} = \frac{1}{3}\pare{5 - \sqrt{7}} \approx 0.7847
 \label{eq:roBSR2}
\end{IEEEeqnarray}

\begin{IEEEproof}  
See Section \ref{sec:PSR}, describing the 3-user PSR scheme.
\end{IEEEproof}
\textcolor{white}{res} \\

\begin{figure}[]
\begin{minipage}{\linewidth}

\centering   


\tikzstyle{every pin}=[font=\footnotesize,pin distance=0.6cm,inner sep=1pt]
\newcommand{\K}{3}
\begin{tikzpicture}
\begin{axis}[
  ymin=0.33,ymax=0.55,xmin=0.5,xmax=3.5,
  xmajorgrids,
   ymajorgrids,
   grid style={dashed, gray!30},
  ylabel style={at={(0.02,0.5)},rotate=-90},
  width=0.48\linewidth, 
   xtick={0,0.5,0.6,0.8,1, 1.263,1.5,1.8,2,2.333,4},
   xticklabels={$0$,$\frac{1}{2}$,$\frac{3}{5}$,$\frac{4}{5}$,$1$,$\frac{24}{19}$,$\frac{3}{2}$,$\frac{9}{5}$,$2$,$\frac{7}{3}$,$4$},
   ytick={0.375, 0.4285,0.5,0.5455},
   yticklabels={$\frac{3}{8}$, $\frac{3}{7}$,$\frac{1}{2}$,$\frac{6}{11}$},
   extra y ticks={0.4444,0.387, 0.333},
   extra y tick labels={$\frac{4}{9}$,$\frac{12}{31}$, $\frac{1}{3}$},
   x tick label style={xtick pos = left}, 
   y tick label style={ytick pos = left, yticklabel pos=left}, 
   extra y tick style={ytick pos = right, yticklabel pos=right},
  font=\footnotesize,
  xlabel=$\rho$,
  ylabel=$\djin$,
  legend style={at={(axis cs: 2,0.35)},anchor=south west}],
  ]
  
  
  
  \addplot[plotoptsUp,domain=1/(\K-1):\K/(\K*\K-\K-1)]{(\K-1)/\K*x};
  \addlegendentry{Outer bound};
  \addplot[forget plot, plotoptsUp,domain=\K/(\K*\K-\K-1):18/22]{(\K-1)/\K*x};

  
  \addplot[forget plot, plotoptsUp,domain=18/22:\K+0.5]{6/11};

      

  \poligon{(axis cs:0.5,0.333) -- (axis cs:0.6,0.375) -- (axis cs:0.7814,0.375) -- (axis cs:0.7814,0.333) -- (axis cs:0.5,0.333)} 
  
  \newcommand{\KK}{3}
  \addplot[forget plot,plotoptsPropIn,domain=1/2:\KK/(\KK*\KK-\KK-1)]{\KK/\K*x/(x+1)}; \label{plot:propin}
  \addplot[refstyle={plot:propin},domain=\KK/(\KK*\KK-\KK-1):0.781,]{\KK/\K*\KK/(\KK*\KK-1)};

  
  
   \poligon{(axis cs:0.7814,0.333) -- (axis cs:0.7814,0.375) -- (axis cs:0.81,0.3870) -- (axis cs:1,0.3870) -- (axis cs:0.8611,0.333) -- (axis cs: 0.7814,0.333) } 
  
    \linianf{0.781}{0.375}{0.781}{0.333} 

  \addplot[refstyle={plot:propin},domain=0.7797:0.7847]{(2*x*x)/(5*x*x-10*x+8)};
  \addplot[refstyle={plot:propin},domain=0.7847:4/5]{6*x/(3*x+10)};
  \addplot[refstyle={plot:propin},domain=4/5:24/19]{12/31};

  

  \poligon{(axis cs:1.263,0.3871) -- (axis cs:1.384,0.3871) -- (axis cs:1.8,0.4286) -- (axis cs:1.5,0.4286) -- (axis cs:1.263,0.3871) } 
  
  \newcommand{\G}{3}
  \addplot[forget plot, refstyle={plot:propin},domain=24/19:3/2]{\G/\K*x/(x+\G-1)};
  \addplot[forget plot, refstyle={plot:propin},domain=3/2:9/5]{3/7};

  
  \addplot[refstyle={plot:propin},domain=\K:\K+0.5]{2/(\K+1)};
  \addplot[plotoptsPropIn,domain=9/5:\K]{2/\K*x/(x+1)};  
  \addlegendentry{New inner bound};
  
  
  
  \addplot[forget plot,plotoptsPrevIn, domain=2:3.5]{1/2}; \label{plot:previn}
  \addplot[refstyle={plot:previn}, domain=18/13:1.98]{2/\K*x/(x+1)};
  \addplot[refstyle={plot:previn}, domain=1:18/13]{12/31};
  \addplot[refstyle={plot:previn}, domain=31/36:1]{12/31*x};
  \addplot[refstyle={plot:previn}, domain=1/\K:31/36]{1/\K};
  \addplot[plotoptsPrevIn, domain=0:1/\K]{x};
  \addlegendentry{Current inner bound}; 
  
  \addplot[forget plot,thick, blue, mark=*,mark size = 2pt, mark options={fill=white,solid},domain=1.999:2]{4/9};
  \addplot[forget plot, thick,blue,mark=*,mark size = 2pt, mark options={ solid,fill=blue},domain=2:2.001]{1/2}; 
  
  \addplot[plotoptsNoCSIT, domain=1/(\K-1):5]{1/\K};
  \addlegendentry{TDMA};

  \textboxx{0.93}{0.40}{\scriptsize RIA $(L=3)$}
  \linia{0.75}{0.395}{0.65}{0.369}
  
  \textboxx{1.14}{0.352}{\scriptsize PSR}
  \linia{1.11}{0.358}{0.88}{0.375}
  
  \textboxx{1.67}{0.465}{\scriptsize TG $(G_2=3)$}
  \linia{1.65}{0.46}{1.52}{0.41}

  \textboxx{3}{0.465}{\scriptsize TG $(G_2=2)$}

\end{axis}
\end{tikzpicture}

\caption{Normalized DoF inner and outer bounds per user for the 3-user MIMO IC with delayed CSIT. Shaded regions identify
where proposed inner bounds improve state-of-the-art.} 
\label{fig:mainResults3}

\end{minipage}
\end{figure}
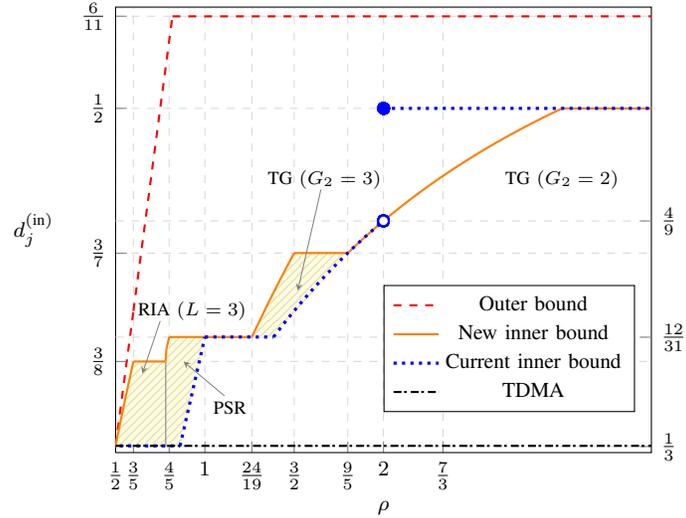

\begin{minipage}{\linewidth}
\begin{theorem}[DoF Inner bound for $K$ users]
For the $K$-user MIMO IC with delayed CSIT and antenna ratio $\rho$, the following DoF per user can be achieved:
 \label{th:innerbound} 
\end{theorem} 
\def\arraystretch{2.8}
\begin{center}
\begin{tabular}{C !{\vrule width 2pt} C !{\vrule width 2pt} c}
\djin & \rho & Scheme \\[1mm]
\noalign{\hrule height 1.5pt} 
      \dfrac{\rho}{\rho+1} & \left(\dfrac{1}{K}, \roA(K)\right) &  \\[2.5mm] 
      \cline{1-2} 
      \displaystyle \dfrac{1}{K} \max \left(
      \dfrac{\lambda^2}{\lambda^2-1}, (\lambda-1)\dfrac{\rho}{\rho+1} \right)  &  
      \left[ \roA(\lambda) , \roA(\lambda-1) \right] ,\, \lambda \in \{4\dots K\}	&  RIA  \\[2.5mm] 
      \cline{1-2} 
      \dfrac{9}{8K}& 
      \left( \dfrac{3}{5} , \rox \right] & \\[2.5mm] 
      \noalign{\hrule height 1.5pt} 
      \dfrac{3}{K}\Gamma& 
      \left( \rox , \roy(K) \right] & 3-user PSR \\[2.5mm]
      \noalign{\hrule height 1.5pt} 
      \dfrac{\rho}{\rho+\pare{K-1}} & 
      \left( \roy(K) , \roB(K) \right) &  \\[2.5mm] 
      \cline{1-2} 
      \displaystyle \max \left(\dfrac{1+\alpha(\epsilon) \!\cdot \! \pare{\epsilon-1}}{K+\pare{1+\epsilon \!\cdot \! \parep{\epsilon-2}} \!\cdot \!\binom{K}{\epsilon}},
      \dfrac{\epsilon-1}{K}\dfrac{\rho}{\rho+\pare{\epsilon-2}} \right)  &	 
      \left[ \roB(\epsilon) , \roB(\epsilon-1) \right],\, \epsilon \in \{3\dots K\} 	& TG   \\[2.5mm] 
      \cline{1-2} 
      \dfrac{2}{K+1} & 
      \left( K, \infty \right) &
\end{tabular} 
\end{center}
\end{minipage}
\textcolor{white}{res} \\[3mm]
where $\roA(\lambda) = \frac{\lambda}{\lambda^2-\lambda-1}$, $\rox= \sqrt{249}-15 \approx 0.7797$,
$\Gamma$ denotes the DoF achieved for the 3-user MIMO IC, and stated in \mbox{Theorem \ref{th:innerbound3}}, applicable to the $K$-user case by means of time-sharing arguments, $\roy(K)=\frac{36(K-1)}{31K-36}$, $\alpha(\epsilon) = \binom{K-1}{\epsilon-1}$, and $\roB(\epsilon) = \frac{1+\alpha(\epsilon) \cdot \pare{\epsilon-1}}{1+\alpha(\epsilon)\cdot \pare{\epsilon-2}}$. Note that $\roA(3)=\frac{3}{5}$, and $\roB(2)=K$, both representing the extremal values of the range of application for the RIA and TG schemes, respectively. \\

\begin{figure}[]
\begin{minipage}{\linewidth}
\centering

\tikzstyle{every pin}=[font=\footnotesize,pin distance=0.6cm,inner sep=1pt]

\begin{tikzpicture}
\begin{axis}[
  ymin=0.16,ymax=0.42,xmin=0.2,xmax=6.5,
  xmajorgrids,
   ymajorgrids,
   grid style={dashed, gray!30},
  ylabel style={at={(0.02,0.5)},rotate=-90},
  width=0.48\linewidth, 
   xtick={0,0.2,0.7847,1.22,3,5,6},
   xticklabels={$0$,$\frac{1}{5}$,$\rox$,$\roy(6)\!$,$3$,$5$,$6$},
   extra x ticks={},
   extra x tick labels={},
   ytick={0.1666, 0.2006,0.2442,0.4082},
   yticklabels={$\frac{1}{6}$, $\frac{875}{4362}$,$\frac{21}{86}$,$\frac{20}{49}$},
   extra y ticks={0.2857,0.1935,0.2222, 0.3165},
   extra y tick labels={$\frac{2}{7}$,$\frac{6}{31}$,$\frac{2}{9}$, $\frac{25}{79}$},
   x tick label style={xtick pos = left}, 
   y tick label style={ytick pos = left, yticklabel pos=left}, 
   extra x tick style={xtick pos = right, xticklabel pos=right},
   extra y tick style={ytick pos = right, yticklabel pos=right},
  font=\footnotesize,
  xlabel=$\rho$,
  ylabel=$\djin$,
  legend style={at={(axis cs: 1.25,0.4)},anchor=north west}],
  ]
\newcommand{\K}{6}
  
  \poligon{(axis cs:0.2,0.1666) -- (axis cs:0.79,0.1666) -- (axis cs:0.79,0.1875) -- (axis cs:0.6,0.1875) -- (axis cs:0.55,0.17778) -- (axis cs:0.37,0.17778) -- (axis cs:0.35,0.17361) -- (axis cs:0.27,0.17361) -- (axis cs:0.25,0.17143)
  -- (axis cs:0.21,0.17143) -- (axis cs:0.2,0.1666)} 
   
  \linianf{0.79}{0.1666}{0.79}{0.1875}
  
  \poligon{(axis cs:0.79,0.1666) -- (axis cs:0.8444,0.1666) -- (axis cs:0.965,0.1935) -- (axis cs:0.79,0.1935) -- (axis cs:0.79,0.1666)} 
  \poligon{(axis cs:1.27,0.2) -- (axis cs:1.52,0.2) -- (axis cs:2,0.2222) -- (axis cs:5,0.2222) -- 
  (axis cs:5,0.27778)  -- (axis cs:4,0.2667) -- (axis cs:3,0.25) -- (axis cs:2.73,0.2441)  -- (axis cs:1.91,0.2441) -- (axis cs:1.56,0.21986) 
  -- (axis cs:1.48,0.21986) -- (axis cs:1.34,0.20588) -- (axis cs:1.27,0.2)} 
  

  \pgfplotstableread[col sep=tab]{DoFsota.txt} \dofsota
  \pgfplotstableread[col sep=tab]{DoFs.txt} \dof

  
  \addplot[plotoptsUp] 
  table[x =ro, y = dout]   from \dofsota ;  
  
  \addlegendentry{Outer bound};

  
  \addplot[plotoptsPropIn] 
  table[x =ro, y = dprop]   from \dof ; 
  \addlegendentryexpanded{New inner bounds};

%
%
%
%
%

    \textboxx{0.9}{0.23}{3-user PSR}
   \linia{0.85}{0.222}{0.85}{0.185}

  \linia{1.2}{0.175}{0.68}{0.175}
   \textboxx{2.23}{0.175}{RIA $(L=3,\dots,6)$}

   \linia{3.6}{0.2852}{3.3}{0.24}
   \textboxx{3.9}{0.291}{TG $(G_2=2,\dots,6)$}
   

  
  \addplot[plotoptsPrevIn]   
  table[forget plot,x =ro, y = dprev,restrict x to domain=0:4.99]   from \dofsota ;
  \addplot[plotoptsPrevIn] 
  table[x =ro, y = dprev,restrict x to domain=5.01:6.5]   from \dofsota ; 
  \addlegendentry{Current inner bound};
  
  \addplot[forget plot,thin, blue, mark=*,mark size = 2pt, mark options={fill=white,solid},domain=4.99:5]{2/9};
  \addplot[forget plot, thin,blue,mark=*,mark size = 2pt, mark options={ solid,fill=blue},domain=5:5.01]{25/79};

  \addplot[plotoptsNoCSIT, domain=1/(\K-1):\K+0.5]{1/\K};
  \addlegendentry{TDMA};
  
  
%
%
%
%

\end{axis}
\end{tikzpicture}

\caption{Normalized DoF inner and outer bounds per user for the 6-user MIMO IC with delayed CSIT. Shaded regions identify
where proposed inner bounds improve state-of-the-art.}
\label{fig:mainResults} 

\end{minipage}
\end{figure}
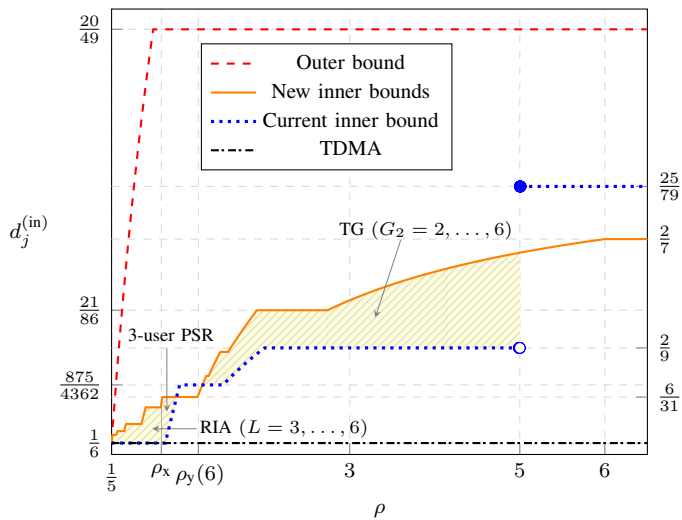   

\begin{IEEEproof}  
Each DoF value is achieved by means of the precoding scheme indicated in the last column. The RIA scheme gets the best performance for $\rho < \rox$, and it is described in Section \ref{sec:innerboundRIA}. When $\rho$ is close to one ($M\approx N$), the 3-user PSR scheme combined with time-sharing performs the best. This scheme is described in Section \ref{sec:PSR}. Finally, the TG scheme addressed in Section \ref{sec:innerboundTG} corresponds to the cases $\rho > \roy(K)$. 
\end{IEEEproof}
\vskip 5mm


Combining \mbox{Theorems \ref{th:innerbound3} and \ref{th:innerbound}}, the inner and outer bound DoF per user for $K=3$ and $K=6$ users are summarized in \mbox{Fig. \ref{fig:mainResults3}} and \ref{fig:mainResults}, respectively. They are represented for $\rho > \frac{1}{K-1}$, since otherwise the DoF outer bound is attained without the need of CSIT, i.e. TDMA, see e.g. \cite{AlignmentChainsTrans}.

Current inner bound curves are constructed by using three different transmission strategies, yielding the best known DoF for each antenna setting. First, the PSR scheme in \cite{Abdoli_IC} for the $K$-user SISO IC may be trivially extended for $M\neq N$ by turning off the additional antennas, and scaling all the parameters by a factor ${\Min{M,N}}$. Second, the scheme for the 2-user MIMO IC in \cite{Vaze2IC} with delayed CSIT is considered, where the equivalent DoF are multiplied by a factor $\frac{2}{K}$, see (\ref{sec:timesharing}). And finally, the work of Hao et al. \cite{Hao_BSR_MISOIC} appeared during the writing of this work has provided new results for the case $M\geq K-1$, $N=1$. Although not explicitly stated in the paper, since all the schemes on delayed CSIT scale, it is assumed that \cite{Hao_BSR_MISOIC} is applicable to the case $\rho=K-1$, and to all cases with $\rho \geq K-1$.

\begin{figure}[h]
  \centering   

\begin{tikzpicture}
\begin{axis}[
  ymin=0,ymax=0.7,xmin=0,xmax=0.6,
  xmajorgrids,
   ymajorgrids,
   grid style={dashed, gray!30},
  ylabel style={at={(0.02,0.5)},rotate=-90},
  width=0.48\linewidth, 
   xtick={0.5,0.333,0.25,0.2,0.1667,0.6},
   xticklabels={$\frac{1}{2}$,$\frac{1}{3}$,$\frac{1}{4}$,$\frac{1}{5}$,$\frac{1}{6}$,$\frac{3}{5}$},
   x tick label style={xtick pos = left}, 
   y tick label style={ytick pos = left, yticklabel pos=left}, 
  font=\footnotesize,
  xlabel=$\rho$,
  ylabel=$\text{gap}$,
  legend pos = north west,
  cycle multi list={ solid,dashed \nextlist
    {red,mark=square,mark size = 2pt},{blue,mark=diamond},{orange,mark=o},{magenta,mark=triangle},{black,mark=triangle,mark options={rotate=180,solid}} },    
    every axis plot/.append style={mark size = 2.2pt,mark repeat={10},mark options={solid}},
  ]

\pgfplotstableread[col sep=tab]{zoom.txt} \gaps
    
    
    \foreach \k in {3,4,...,7} {
      \addplot table[x =ro, y = Kprev\k] from \gaps ;
      \addlegendentryexpanded{$K=\k$};
  }
  
   
   \foreach \k in {3,4,...,7} {
      \addplot table[x =ro, y = Knew\k] from \gaps ;
  }
  \end{axis}
\end{tikzpicture}

\caption{Relative gap for $0 \leq \rho \leq \frac{3}{5}$. The relative gap represents the relative distance from previous (dashed) and new (solid) inner bounds to the best known outer bounds.}
\label{fig:zoom} 
\end{figure}
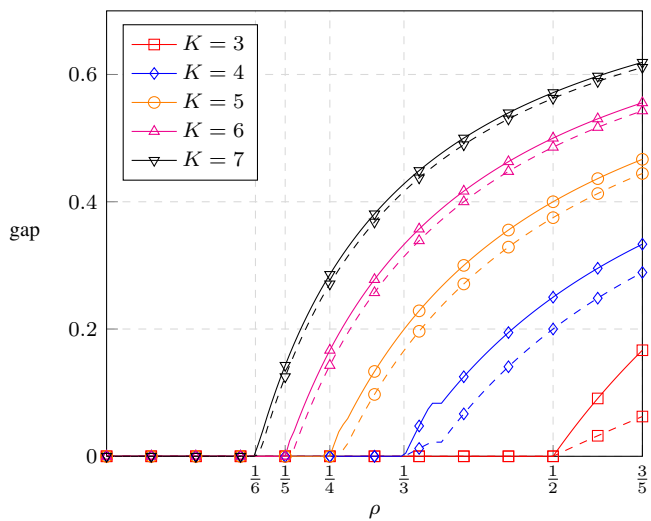

No claim of optimality for the proposed inner bounds is stated, while it is worth pointing out that they outperform current inner bounds for many antenna settings. Moreover, for the region $\frac{1}{K-1}<\rho<\frac{K}{K^2-K-1}$, the RIA scheme gets close to the best known DoF outer bound. To emphasize this, the relative gap for $K=3,\dots,7$, $\rho<\frac35$ is depicted in \mbox{Fig. \ref{fig:zoom}}, defined as:
\begin{IEEEeqnarray*}{c}
 \text{gap} = \frac{\djout-\djin}{\djout}.
\end{IEEEeqnarray*}
The figure shows that for $\rho < \frac{1}{K-1}$ the DoF outer bound is attained. On the other hand, for the region $\frac{1}{K-1}<\rho<\frac{K}{K^2-K-1}$ the new inner bounds provide a much smaller relative gap as compared to the previous inner bounds. And finally, for $\frac{K}{K^2-K-1}<\rho <\frac35$ the relative gap is significant for both previous and new inner bounds, which claims for the research of new and tighter outer bounds.
 
{One may ask which of the previous results is applicable in case there is delayed CSIT, but the channel remains constant. In other words, are previous results applicable without assuming time-varying channels? In this regard, the following is stated:}
\vskip 5mm

\begin{minipage}{\linewidth}
\begin{theorem}[DoF Inner bound with delayed CSIT and constant channels]
All inner bounds proposed in Theorem \ref{th:innerbound} apply for the $K$-user MIMO IC with delayed CSIT, constant channels, and antenna ratio $\rho$.
 \label{th:constant} 
\end{theorem}
\end{minipage}
\textcolor{white}{res} 
\begin{IEEEproof}  
See Section \ref{sec:constant}.
\end{IEEEproof}
\vskip 5mm


\begin{figure}[h]
\begin{minipage}[]{1\linewidth}
  \centering  
  \centerline{\includegraphics[width=0.3\linewidth]{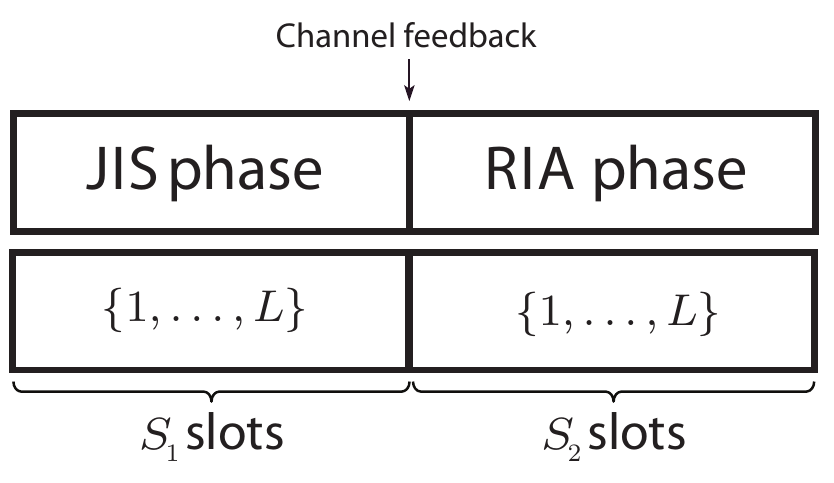}}
\end{minipage}  
\caption{Transmission frame for the RIA scheme. $P=2$ single-round phases, where only $L$ out of the total $K$ users are served, i.e. $\Ac^{(1)}=\Ac^{(2)}=\{1,\dots,L\}$. By time-sharing concepts, the rest of users are considered on other frames.}  
\label{fig:frameRIA} 
\end{figure}
 
\section{RIA scheme $(M < N)$}
\label{sec:innerboundRIA} 
%
%
%
This two-phase scheme is general for the $K$-user MIMO case, and proves Theorem \ref{th:innerbound} for $\rho<\rox$. Next section gives an intuition behind this transmission strategy. Then, each of the two phases is built, and finally we present the optimization problem that provides the optimal system parameters for any antenna setting and number of users.

\subsection{Overview of the precoding strategy} 
\label{sec:overview}

The transmission frame is depicted in \mbox{Fig. \ref{fig:frameRIA}}, where in both phases only $L \leq K$ users are scheduled for the communication, with $L\in \{3,\dots,K\}$.

The two phases are addressed in Sections \ref{sec:ISphase} and \ref{sec:RIAphaseRIA}, and denoted as the joint interference sensing (JIS) phase, and the retrospective IA (RIA) phase, respectively. Delayed-CSIT precoding and redundancy transmission constitute the two main ingredients. All $L$ users considered are active during two single-round phases, i.e:
\begin{IEEEeqnarray}{c}
R_1=R_2=1, \\[2mm]
 G_1 = G_2 = L.
\end{IEEEeqnarray}
{During the JIS phase the transmitted signals are precoded with coefficients agreed before the communication. The objective is that each receiver senses the interference, with the objective of being used during the second phase. Thanks to the channel feedback, at beginning of the RIA phase each transmitter is able to reconstruct the interference terms generated at the non-intended receivers in the previous phase. Then, the transmitted signals delivered during the second phase are designed to be aligned with the interference generated during the first phase, i.e. such that do not cause additional interference.}

Next two sections describe the transmission scheme for a particular value of $L$. 
The methodology used to derive the optimal value of $L$, as well as the optimal system parameters for each antenna setting $\rho$ is addressed in \mbox{Section \ref{sec:RIAopt}}. Table \ref{tab:summary_A} shows the optimal system parameters for a given value of $L$, entailing two different antenna setting regimes: ${\AI=\{\frac{1}{K-1} < \rho \leq \roA(L) \}}$ and ${\AII=\{\roA(L) < \rho \leq 1 \}}$. Note that for the regime $\AII$ the achieved DoF are constant with respect to $M$, and equal to the achievable DoF for $\rho=\roA(L)$. Actually, this simply evidences that if a DoF value can be attained for $\rho=\roA(L)$, it is also achievable for $\rho>\roA(L)$. In particular, those cases may be tackled by scaling equally all the parameters and turning off enough transmit antennas to obtain the desired antenna ratio\footnote{This methodology might not be possible if parameters are limited to some value for the sake of e.g. low complexity or communication delay, as in \mbox{Section \ref{sec:practical}}.}. Consequently, without loss of generalization, in what follows regime $\AI$ is detailed only.

\begin{table}[h]
\caption{System parameters for RIA scheme as a function of $\rho$}
\begin{center}
\begin{tabular}{c | c | c | c | c}
  &  $b$  & $S_1$ & $S_2$ & $\djin$  \\[1mm]
\cline{1-5} \bigstrut 
$\AI=\{\frac{1}{2} < \rho \leq \roA(L)\} $ & $MN $ & $N$ & $M$ & $\frac{L}{K}\frac{\rho}{\rho+1}$ \\
\cline{1-5} \bigstrut
 $\AII=\{ \roA(L) < \rho < 1 \} $ & $LN$ & $L^2-L-1$ & $L$ & $\frac{L}{K}\frac{L}{L^2-1}$ \\
\end{tabular} 
\end{center}
\label{tab:summary_A}
\end{table} 

\subsection{Joint interference sensing phase} 
\label{sec:ISphase}
The first phase lasts for $S_1$ slots with all transmitters active. Since there is no CSI available at the transmitters, generic full-rank precoding matrices $\Vn_i^{(1)} \Cmat{MS_1}{b}$ selected from a predetermined dictionary are agreed by all nodes. As specified in \refbox{Table}{tab:summary_A}, $b=MN$, $S_1 = N$. Then, each receiver obtains $NS_1 = N^2$ observations, which are processed and write as
\begin{IEEEeqnarray}{l}
\zn_j^{\left(1\right)} =  \Un_j^{(1)} \Hn_{j,j}^{\left(1\right)} 
 \Vn_j^{\left(1\right)}  \mathbf{x}_j  + \Un_j^{(1)}
 \left[
      \Hn_{j,\Ic^j_{1}}^{\left(1\right)}   \Vn_{\Ic^j_{1}}^{\left(1\right)} \dots \Hn_{j,\Ic^j_{L-1}}^{\left(1\right)}   \Vn_{\Ic^j_{L-1}}^{\left(1\right)}
 \right] 
 \begin{bmatrix}
      \xn_{\Ic^j_{1}} \\[-2mm] \vdots \\[-2mm] \xn_{\Ic^j_{G-1}}
 \end{bmatrix}
   ,
 \label{eq:scA:1phase}
\end{IEEEeqnarray} 
where $\Ic^j = \{1,\dots,L\} \backslash \{j\} $, and $\Ic^j_k$ is the $k$th index of the set $\Ic^j$, and as explained above the noise term is omitted since we focus on DoF analysis. 

{We impose by design that the parameters satisfy $NS_1 > (L-2)b$, thus there is some redundancy: the receiver has more observations than variables corresponding to interference. This redundancy can be exploited in pursuit of partial interference nulling, i.e. projecting the received signals onto subspaces where the desired signals are interfered by the symbols of a single user. In this regard,} let define the receiving filter ${\Un_{j}^{(1)} \Cmat{\varphi_0}{NS_1}}$, ${\forall i \neq j}$, with 
\begin{IEEEeqnarray*}{c}
\varphi_0=(L-1) \varphi_1, \\ {\varphi_1 =  NS_1-(L-2)b} = N\pare{N- (L-2)M},
\end{IEEEeqnarray*}
which consists of the composition of $L-1$ linear filters ${\Un_{j,i}^{(1)} \Cmat{\varphi_1}{NS_1}}, i\neq j$, defined such that
\begin{IEEEeqnarray}{c}
\begin{matrix}
\Un_{j,i}^{(1)} \Hn_{j,k}^{\left(1\right)}   \Vn_{k}^{\left(1\right)} = \0,\, k \neq \{i,j\}  \\
\Un_{j,i}^{(1)} \Hn_{j,i}^{\left(1\right)}   \Vn_{i}^{\left(1\right)} \neq \0, \\[1.5mm]
\Un_j^{(1)} = \stackp{\Un_{j,\Ic^j_{1}}^{(1)},\dots,\Un_{j,\Ic^j_{L-1}}^{(1)}}, \\[2.5mm]
\Un_j^{(1)} \Big[
      \Hn_{j,\Ic^j_{1}}^{\left(1\right)}   \Vn_{\Ic^j_{1}}^{\left(1\right)} \dots \Hn_{j,\Ic^j_{L-1}}^{\left(1\right)}   \Vn_{\Ic^j_{L-1}}^{\left(1\right)}
 \Big] 
 = \bdiag{\Tn_{j,\Ic^j_{1}},\dots,\Tn_{j,\Ic^j_{L-1}}},
\end{matrix} \label{eq:scA:PIC_ISph}  \\
{\Tn_{j,i} = \Un_{j,i}^{(1)} \Hn_{j,i}^{\left(1\right)} \Vn_{i}^{\left(1\right)} \Cmat{\varphi_1}{b}},i\neq j
\label{eq:scA:OHI} 
\end{IEEEeqnarray}
where $ \Tn_{j,i}$ is the residual interference from $\BS{i}$ after applying the linear filter $\Un_{j,i}^{(1)}$,
i.e. this processing together with the transmitted redundancy allows uncoupling the interference from the different sources at $\UE{j}$. 
Now, let define for each $i\neq j$ the subspace 
\begin{IEEEeqnarray}{c}
 \Tc_{j,i} = \RSpan{\Tn_{j,i}}.
 \label{eq:Tji}
\end{IEEEeqnarray}
Those subspaces represent the {\it overheard interference} the signals of the second phase can be aligned with. Notice that they can be constructed using only delayed CSIT, thus transmitters will be able to construct them at the beginning of the second phase. Finally, for the sake of reader's understanding, let write the processed signals in (\ref{eq:scA:1phase}) by applying the design for $\Un_j^{(1)}$ in (\ref{eq:scA:PIC_ISph}):
\begin{IEEEeqnarray}{l}
\zn_j^{\left(1\right)} =  \Un_j^{(1)} \Hn_{j,j}^{\left(1\right)} 
 \Vn_j^{\left(1\right)}  \mathbf{x}_j  + 
 \begin{bmatrix}
      \Tn_{j,\Ic^j_{1}} \, \xn_{\Ic^j_{1}} \\ \vdots \\ \Tn_{j,\Ic^j_{G-1}} \, \xn_{\Ic^j_{G-1}}
 \end{bmatrix}.
 \label{eq:scA:1phaseProc}
\end{IEEEeqnarray}

\vskip 3mm

\subsection{Retrospective Interference Alignment phase} 
\label{sec:RIAphaseRIA}

The second phase lasts for $S_2 = M$ slots where the precoding matrix for $\Tx{i}$ is designed to align the generated interference with the overheard interference at {\it all} non-intended receivers. In other words, each receiver should be able to remove the interference generated by $\Vn_{i}^{(2)}$ using the overheard interference from the JIS phase, see (\ref{eq:Tji}). Then, they are designed to satisfy the following set of constraints:
\begin{IEEEeqnarray}{c}
\RSpan{ 
      \Hn_{k,i}^{(2)} \Vn_{i}^{(2)} } 
      \subseteq 
      \Tc_{k,i}, \, \forall k \neq i \label{eq:scA:RIA_cond}.
\end{IEEEeqnarray}
An easy way to ensure this without using full CSIT is to set
\begin{IEEEeqnarray}{c}
 \mathbf{V}_{i}^{(2)}  =  \boldsymbol{\Sigma}_i^{(2)} \Tn_{i}^{(2)} \label{eq:scA:RIA_simple},  \\[2mm]
 \RSpan{\Tn_i^{(2)}} = \Tc_i^{(2)} =  \bigcap_{k \neq i} \Tc_{k,i}  \label{eq:scA:Tdef},
\end{IEEEeqnarray}
where $\boldsymbol{\Sigma}_i^{(2)} \Cmat{MS_2}{\varphi_2}$ is some arbitrary full rank matrix ensuring the transmit power constraint, and $\Tn_i^{(2)} \Cmat{\varphi_2}{b}$ is some arbitrary matrix whose rows span the intersection subspace $\Tc_i^{(2)}$ of dimension 
\begin{IEEEeqnarray}{r c l}
\varphi_2 & \, &= b - (L-1)(b-\varphi_1) \label{eq:scA:dimIntersection} \\
& \, &= N \pare{(L-1)N -L(L-2)M}, \nonumber
\end{IEEEeqnarray}
derived using identity (\ref{eq:intersection}). The received signals along the whole communication at each receiver, can be more easily understood by writing the $j$th signal space matrix:
\begin{IEEEeqnarray*}{c}
\def\arraystretch{1.7}
 \Gn_j = \left[ \,
\begin{matrix}
   \Un_{j,\Ic^j_{1}} \Hn_{j,j}^{\left(1\right)}   \Vn_{j}^{\left(1\right)} & \Tn_{j,\Ic^j_{1}} &     \cdots & \0          \\ 
   \vdots & \vdots & \ddots & \vdots \\
   \Un_{j,\Ic^j_{G-1}} \Hn_{j,j}^{\left(1\right)}   \Vn_{j}^{\left(1\right)} &      \0    & \cdots  & \Tn_{j,\Ic^j_{G-1}}     
   \\[1mm]  \cdashline{1-4}[3pt/2pt]  \\[-6.5mm] 
  \Hn_{j,j}^{(2)} \Vn_j^{(2)} & \Hn_{j,\Ic^j_{1}}^{(2)} \Vn_{\Ic^j_{1}}^{(2)} & \cdots & \Hn_{j,\Ic^j_{G-1}}^{(2)} \Vn_{\Ic^j_{G-1}}^{(2)}
\end{matrix} \, \right] ,
\label{eq:scA:matrixG}
\end{IEEEeqnarray*}
where the dotted lines separate the blocks rows corresponding to each of the two phases. {Note that combination of processed signals may be interpreted as row operations on the signal space matrix.}
Since precoding matrices satisfy conditions in (\ref{eq:scA:RIA_cond}), each interference term generated during the second phase is aligned with one of the overheard interference terms of the first phase.  Therefore, all the second phase interference can be removed, and $N$ LCs of desired symbols free of interference are retrieved at each receiver per time slot, i.e. $NS_2 = MN=b $ LCs after all. In the next section, the constraints to be satisfied by all parameters for each antenna setting will be presented, including that all such $b$ LCs are linearly independent, and thus all desired symbols can be linearly decoded.

Finally, after explaining this precoding scheme we are able to highlight the main difference of the IC w.r.t. the BC. In this case, each transmitter has only access to its own symbols, thus can only reconstruct part of the overheard interference. Consequently, the interference can only be aligned individually, i.e. two users cannot align their signals simultaneously at one receiver with the signals of one slot, since the transmitted signals travel through different channels. This is why a partial interference nulling is applied to the first phase received signals by means of the processing filter $\Un_j^{(1)}$, such that only one interference term affects the desired signals on the processed signal space. {In terms of the signal space matrix, this means that block columns corresponding to interference should have at most one non-zero element per block row.}

\subsection{System parameters optimization}
\label{sec:RIAopt} 
 
Optimal system parameters for each antenna setting and number of users are derived next.
First, the optimal value of $L$ can be found by exhaustive evaluation of the expressions in \mbox{Theorem \ref{th:innerbound}}. Since for high values of $K$ there will be many regions, the Algorithm 1 in the next page is provided to alleviate the search for the optimal $L$ to only two candidates. The motivation behind each of its different steps is next explained. 

First, the real number $x$ is the positive solution of inverting the definition of $\roA(L)$, defined in Theorem \ref{th:innerbound}. Then, since the inner bound is a piecewise function, $x$ represents the value of $L$ between two steps. For this reason, using the ceil and floor functions the two closest integers are selected as candidates, evaluating the achievable DoF for each of them. Finally, the best integer value $L$ is chosen taking into account the extreme cases.

\begin{table}[h!]
\begin{center}
 \begin{tabular}{l l}
 \toprule
 \multicolumn{2}{c}{{\bf Algorithm 1}: $L$ solver} \\
 \hline  \bigstrut
{\bf Step 1}: & $ x := \frac{1}{2} \pare{ \pare{1+\rho^{-1}} + \sqrt{ \pare{1+\rho^{-1}}^2 + 4}}$\\
{\bf Step 2}: & $ y := \floor{x} \frac{\rho}{\rho+1}$,  $z := \frac{1}{K}\frac{\left\lceil x \right\rceil^2}{\ceil{x}^2-1}$ \\[3mm]
{\bf Step 3}: & $L(\rho) = \left\{
\begin{matrix} 
      K & x \geq K \\
      3 & x \leq 3 \\
      \floor{x} & 3 < x < K, \, y > z \\
      \ceil{x} & \text{otherwise}
     \end{matrix} \right.$\\
\hline
\end{tabular}
\end{center}
\end{table}

Assuming a particular value for $L$, we formulate the following DoF optimization problem:
\begin{subequations}
\begin{IEEEeqnarray}{c c l}
 \mathcal{P}_1 : & \underset{ \left\{b,S_1,S_2 \right\} \in \enters}{\text{maximize}} \quad  & \frac{L}{KN}\frac{b}{S_1+S_2} \label{eq:objfunc} \\[2mm]
 & s.t. & MS_1 \geq b \label{eq:maxProbRIA1} \\
  &   & NS_1 > (L-2)b \label{eq:maxProbRIA2} \\    
    & & N S_2 \geq b \label{eq:maxProbRIA3}\\
    & & {LM S_2 \geq  b} \label{eq:maxProbRIA3p}\\
    &  & L \varphi_2 \geq b \label{eq:maxProbRIA4},
\end{IEEEeqnarray}
\end{subequations}
with $\varphi_2 = (L-1)NS_1 - L(L-2)b$. This problem provides the optimal values for $b$, $S_1$, and $S_2$ when the RIA scheme is employed. The objective function corresponds to the number of symbols divided by the channel uses, and a factor due to time-sharing and DoF normalization. On the other hand, the following four constraints are introduced to ensure linear feasibility:

\vskip 1.5mm

\subsubsection{Transmit rank during the JIS phase (\ref{eq:maxProbRIA1})}: During the first phase, $MS_1$ linear combinations of the $b$ symbols are transmitted using $M$ antennas, and during $S_1$ slots. Then, for linear decodability of the desired symbols, no more symbols than the number of transmit dimensions can be sent.
\vskip 1.5mm 
 \subsubsection{JIS phase redundancy (\ref{eq:maxProbRIA2})}: After the first phase, the linear filters $\Un_{j,i}\Cmat{\varphi_1}{NS_1}$ in (\ref{eq:scA:PIC_ISph}) are applied assuming some redundancy has been transmitted, with $\varphi_1 = NS_1-(L-2)b$. Then, we force $\varphi_1>0$ or, equivalently, (\ref{eq:maxProbRIA2}).
\vskip 1.5mm
 \subsubsection{Receiver space-time dimensions (\ref{eq:maxProbRIA3})}: Each receiver should have enough space-time dimensions to allocate all the desired and interference signals without space overlapping. First, notice that the interference received during the JIS phase occupies at most $NS_1$ dimensions. This subspace remains the same after the RIA phase, since all the interference generated during the RIA phase is aligned. On the other hand, the desired signals occupy at most $b$ dimensions at each receiver. Hence, we must have
\begin{IEEEeqnarray*}{c}
\underbrace{b_{ }}_{\text{desired dim.}} + \underbrace{NS_1}_{\text{interference dim.}} \leq \underbrace{NS_1 + N S_2}_{\text{total dimensions}}
\label{eq:constraintEspai}
\end{IEEEeqnarray*} 
\vskip 1.5mm

\subsubsection{Rank of desired signals after zero-forcing (\ref{eq:maxProbRIA3p})-(\ref{eq:maxProbRIA4})}: For ease of exposition, the signal space matrix $\Gn_j$ at each receiver is here rewritten: 
\begin{IEEEeqnarray*}{c}
\def\arraystretch{1.7}
\Gn_j = \left[ \,
\begin{matrix}
   \Un_{j,\Ic^j_{1}} \Hn_{j,j}^{\left(1\right)}   \Vn_{j}^{\left(1\right)} & \Tn_{j,\Ic^j_{1}} &     \cdots & \0          \\ 
   \vdots & \vdots & \ddots & \vdots \\
   \Un_{j,\Ic^j_{L-1}} \Hn_{j,j}^{\left(1\right)}   \Vn_{j}^{\left(1\right)} &      \0    & \cdots  & \Tn_{j,\Ic^j_{L-1}}     
   \\[1mm]  \cdashline{1-4}[3pt/2pt]  \\[-6.5mm] 
  \Hn_{j,j}^{(2)} \Vn_j^{(2)} & \Hn_{j,\Ic^j_{1}}^{(2)} \Vn_{\Ic^j_{1}}^{(2)} & \cdots & \Hn_{j,\Ic^j_{L-1}}^{(2)} \Vn_{\Ic^j_{L-1}}^{(2)}
\end{matrix} \, \right]
 \begin{matrix*}[l]
 \updownarrow \varphi_1
 \\[0.3mm]
 \ 
 \\[0.3mm]
 \updownarrow \varphi_1
 \\[1.2mm]
  \updownarrow NS_2
\end{matrix*},
\end{IEEEeqnarray*}
where the block rows corresponding to the first phase have $\varphi_1$ rows each, whereas the block row of the second phase has $NS_2$ rows. Now, recall that the precoding matrices $\Vn_{i}^{(2)}$ lie on a subspace of dimension $t = \Min{MS_2,\varphi_2}<\varphi_1$, see (\ref{eq:scA:RIA_simple})-(\ref{eq:scA:dimIntersection}). Then, if the interference is to be removed, each of the $L-1$ block rows corresponding to the JIS phase must be projected onto the corresponding subspace of dimension $t$ and linearly combined with the block row of the second phase. This is done by means of the linear filter $\Wn_j$, obtaining
\begin{IEEEeqnarray*}{c} 
 \begin{matrix}
  \Rank{\mathbf{W}_{j} \Un_j \mathbf{H}_{j,j} \mathbf{V}_j} = \min\pare{L\cdot\Min{MS_2,\varphi_2},b}.
 \end{matrix}
\end{IEEEeqnarray*}
Since any linear precoding scheme requires $\Rank{\mathbf{{W}}_{j} \Un_j \mathbf{{H}}_{j,j} \mathbf{{V}}_j}\geq b$, this yields to 
\begin{IEEEeqnarray*}{c}
L\cdot \Min{MS_2,\varphi_2}\geq b \Rightarrow 
\begin{cases} 
	LMS_2 \geq b  \\ 
	L \varphi_2 = L \pare{ (L-1)NS_1 - L(L-2)b} \geq b
\end{cases}.
\end{IEEEeqnarray*}

\vskip 2mm

Next, we analytically derive the solution of problem $\mathcal{P}_1$. For any given value of $b$, the objective function in (\ref{eq:objfunc}) is strictly decreasing with $S_1$ and $S_2$, i.e. their optimum values are their minimum feasible values. Therefore, since $S_2$ appears in (\ref{eq:maxProbRIA3}) and (\ref{eq:maxProbRIA3p}) only, its optimum value $S_2^*$ is given by
\begin{IEEEeqnarray*}{c} 
S_2^* =  \Big\lceil b \Max{\frac{1}{N},\frac{1}{ML}} \Big\rceil 
\end{IEEEeqnarray*}
This establishes two regions, with the threshold $\rho = \frac{1}{L}$. However, it can be seen that taking 
$S_2^* = \Big\lceil\frac{b}{ML} \Big\rceil$ and solving the problem produces a DoF value which is always outperformed by taking  
$S_2^* = \Big\lceil \frac{b}{N}\Big\rceil$ and increasing the value of $L$. Hence, we definitely take
\begin{IEEEeqnarray}{c}
S_2^* = \Big\lceil \frac{b}{N} \Big\rceil .
\label{eq:RIA:optS2}
\end{IEEEeqnarray}
On the other hand, the optimum value of $S_1$ is set to satisfy one of the constraints (\ref{eq:maxProbRIA1}), (\ref{eq:maxProbRIA2}), and (\ref{eq:maxProbRIA4}) with equality:
\begin{IEEEeqnarray}{c}
S_1^* \! = \left\lceil \max \!\left( {\frac{b}{M}}, \frac{b+1}{N}, \frac{b}{NL}\pare{L^2-L-1} \right) \right\rceil =
 \left\lceil b \cdot \max \!\left( {\frac{1}{M}}, \frac{1}{NL}\pare{L^2-L-1} \right) \right\rceil.
%
\label{eq:RIA:optS1}
\end{IEEEeqnarray}
While in \mbox{Section \ref{sec:practical}} a maximum-value constraint for $b$ will be included, here the problem is solved for unbounded $b$, i.e. it is simply chosen such that all parameters are integer values. Accordingly, one optimal solution is specified in Table \ref{tab:summary_A}. Note that the threshold $\roA(L) = \frac{L}{L^2-L-1}$ follows from the two possible choices for $S_1^*$, with $b=MN$ or $b=NL$, for each case.


\section{TG scheme $(M>N)$}
\label{sec:innerboundTG} 

The two-phase TG scheme proves Theorem \ref{th:innerbound} for $\rho>\roy(K)$. 
 Next section gives an intuition behind this strategy. Then, each of the two phases is built, and finally we present the optimization problem that provides the optimal system parameters for any antenna setting and number of users. 

\subsection{Overview of the precoding strategy} 
\label{sec:overview}

This approach is designed according to two main ingredients: {\it delayed CSIT precoding and user scheduling}. In contrast to the RIA scheme, now all users are considered in each transmission block ($L=K$), and scheduled through the different rounds. During the first phase, time resources are orthogonally distributed among users, thus $G_1=1$, such that interference can be sensed individually. For this reason, this phase will be labeled as the individual interference sensing (IIS) phase. Notice also that in addition to sensing the interference, this phase provides free of interference observations of the desired signals to each receiver. 

{Each round of the second phase is dedicated to a different group of $G_2$ users, which for simplicity in the notation will be simply denoted as $G$.} The objective is similar to the second phase of the RIA scheme, and for this reason it is also denoted hereafter as the RIA phase. Based on the channel feedback, each active transmitter is able to send LCs of symbols that can be removed at the non-intended active receivers by exploiting the overheard interference from the IIS phase. As an example, the transmission frame for the case $K=4$, $G=3$ is depicted in \mbox{Fig. \ref{fig:frameTG43}}. 

\begin{figure}[h]
\begin{minipage}[]{1\linewidth}
  \centering  
  \centerline{\includegraphics[width=0.4\linewidth]{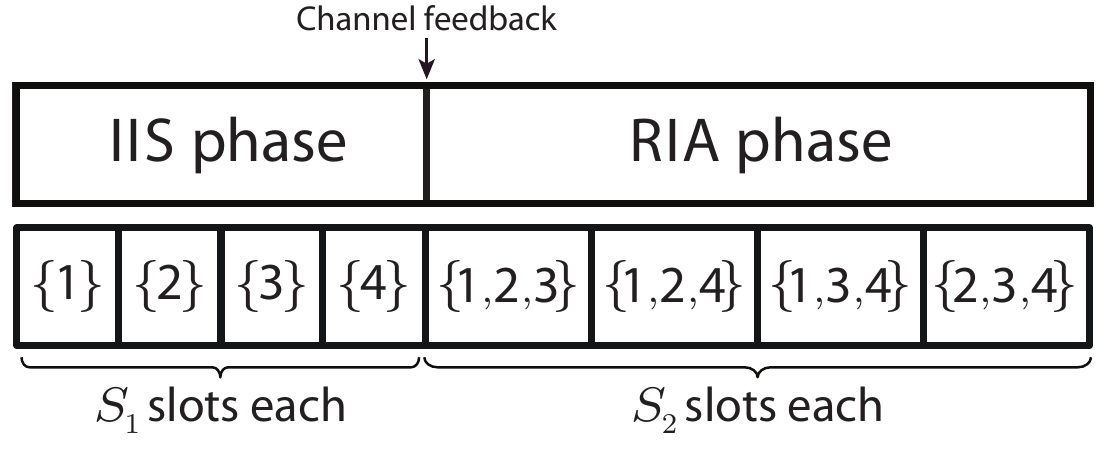}}
\end{minipage}  
\caption{Transmission frame for the TG scheme with $K=4$, $G=3$. Each of the $P=2$ phases has four rounds of $S_1$ and $S_2$ slots each. Active groups $\Ac^{(p,r)}$ are represented for each round of the two phases.}  
\label{fig:frameTG43} 
\end{figure} 

Summarizing, we have
\begin{IEEEeqnarray}{c c c}
R_1 = K, & \quad & R_2 = \binom{K}{G}, \\[2mm]
 G_1 = 1, & \quad & G_2 \triangleq G.
\end{IEEEeqnarray}
{Note that the transmitted signals during each round of the second phase are designed such that they can be removed at the $G-1$ non-intended active receivers. Then, low values of $G$ relax the constraints on the design, i.e. the number of receivers where the transmitted signals should be aligned, but also increase the number of rounds. This is a trade-off that should be balanced by the optimal value of $G$.}
The derivation of the optimal system parameters, as well as $G$ is deferred to \mbox{Section \ref{sec:TGopt}}, and presented in \mbox{Table \ref{tab:summary_B}}. For each value of $G$, it can be seen that there exist two different antenna setting regimes: ${\BI=\{1 < \rho \leq \roB(G) \}}$ and ${\BII=\{ \rho > \roB(G) \}}$, with $\roB(G) = \frac{1+\alpha(G) \cdot  \pare{G-1}}{1+\alpha(G) \cdot \pare{G-2}}$. Following similar arguments as for the RIA scheme, only the case $\BI$ will be addressed in the sequel, and a particular value for $G$ is assumed. Therefore, for ease of notation simply
\begin{IEEEeqnarray*}{c}
\alpha \triangleq \alpha(G) = \binom{K-1}{G-1}
\end{IEEEeqnarray*}
will be used during the following two sections.

\begin{table}[h]
\caption{System parameters for TG scheme as a function of $\rho$}
\begin{center}
\begin{tabular}{c | c | c | c | c | c}
  &  $b$  & $S_1$ & $S_2$ & $\djin$  \\[1mm]
\cline{1-6} \bigstrut 
$\BI=\{1 < \rho \leq \roB(G) \} $ & $\alpha  MN$ & $\alpha  N$ & $M-N$ & $\frac{G}{K}\frac{\rho}{\rho + (G-1)}$ \\
\cline{1-6} \bigstrut
$\BII=\{\rho > \roB(G) \} $ & $(1+\alpha \cdot(G-1))N$ & $1+\alpha \cdot(G-2)$ & $1$ & $\frac{1+\alpha \cdot(G-1)}{K+\pare{G(G-2)+1}\binom{K}{G}} $ \\
\end{tabular} 
\end{center}
\label{tab:summary_B}
\end{table} 
 
\subsection{Individual Interference Sensing phase}
Each $\Tx{i}$ sends linear combinations of its $b = \alpha  MN$ symbols during the $S_1 = \alpha  N$ time slots of the $(1,i)$th round, thus $\Rx{j}$ obtains
\begin{IEEEeqnarray}{c}
\yn_j^{\left(1,i\right)} =  \Hn_{j,i}^{\left(1,i\right)} 
 \Vns{i}{1,i}  \mathbf{x}_i ,
 \label{eq:scD:1phase}
\end{IEEEeqnarray}
where $\Hns{j,r}{1,r} \Cmat{NS_1}{MS_1}$, and the precoding matrices $\Vns{r}{1,r} \Cmat{MS_1}{b}$ are chosen as some generic full-rank matrices. Since no redundancy was transmitted ($b<NS_1$), none per-phase receiving filter is applied, i.e. equivalently we have $\Un_j^{(1)}=\I_{N\tau_1}$. Moreover, similarly to (\ref{eq:scA:OHI}) we define 
\begin{IEEEeqnarray}{c}
\Tn_{j,i} = \Hn_{j,i}^{\left(1,i\right)}  \Vns{i}{1,i} \\[2mm] 
\Tc_{j,i} = \RSpan{\Tn_{j,i}}, 
\end{IEEEeqnarray}
{as the overheard interference generated by $\Tx{i}$ at $\Rx{j}$, with $\Dim{\Tc_{j,i}}=NS_1 = \alpha  N^2$. In contrast to the previous scheme, note that now this term is individually obtained since there is only one active transmitter per round. Specifically, each receiver observes $NS_1 = \alpha  N^2$ linear combinations of the desired symbols, as well as 
$\alpha N^2 (K-1)$ linear combinations of overheard interference, and since $NS_1 < b$, linear decodability is not possible yet.}

\subsection{Retrospective Interference Alignment phase}
\label{sec:innerboundTG:2}
The objective of the RIA phase is to exploit the overheard interference, i.e. the subspaces $\Tc_{j,i}$ available at the non-intended receivers, to construct signals that can be canceled even without knowing the current CSI. The design pursues that for each round $r$ of the second phase, the transmitted signals are aligned at \textit{all} the $G$ receivers in $\Ac^{(2,r)}$. For this reason, the optimal value of $G$ depends on each antenna setting and the total number of users $K$.

According to this objective, the signal transmitted during the $(2,r)$th round by each active transmitter $i\in\Ac^{(2,r)}$ should satisfy the following set of constraints:
\begin{IEEEeqnarray}{c}
\RSpan{ 
      \Hn_{k,i}^{(2,r)} \mathbf{V}_{i}^{(2,r)} } 
      \subseteq 
      \Tc_{k,i}, \forall k \in \Ac^{(2,r)}\backslash \{i\} \label{eq:scB:RIA_cond}.
\end{IEEEeqnarray}
This can be ensured by setting 
\begin{IEEEeqnarray}{c}
\begin{matrix}
  \mathbf{V}_{i}^{(2,r)}  =  \boldsymbol{\Sigma}_i^{(2,r)} \Tn_{i}^{(2,r)} \label{eq:scB:RIA_simple}, \\[2mm]
 \displaystyle \RSpan{\Tn_i^{(2,r)}} = \Tc_i^{(2,r)} =  \!\! \bigcap_{{k\in \Ac^{(2,r)}\backslash \{i\} }}  \!\!\!\!\Tc_{k,i}   , 
\end{matrix}
\label{eq:scB:Tdef} 
\end{IEEEeqnarray}
where $\boldsymbol{\Sigma}_i^{(2,r)} \Cmat{MS_2}{\varphi}$ is some arbitrary full rank matrix ensuring the transmit power constraint, and $\Tn_i^{(2,r)} \Cmat{\varphi}{b}$ is a matrix whose rows lie on the intersection subspace of dimension
\begin{IEEEeqnarray}{r c l}
 \varphi \, & \, & = (G-1)NS_1 - (G-2)b \label{eq:scB:dimIntersection} \\ \nonumber
& \, & = \alpha  N \pare{N(G-1) - (G-2)M}.
\end{IEEEeqnarray}
{The main difference between the second phase of this scheme w.r.t. to the second phase of the RIA scheme is that the transmitted signals should be removable only at the non-intended active receivers, instead that at all receivers. This can be seen by comparing (\ref{eq:scA:Tdef}) with (\ref{eq:scB:Tdef}): instead of the intersection of all but one subspaces, during each round each transmitter sends signals that lie on the intersection of $G-1$ subspaces.}
In order to illustrate how the signals are received and aligned, the signal space matrix at each receiver for the case $K=4$, $G=3$ is next shown:
\begin{IEEEeqnarray}{c}
 \def\arraystretch{1.7}
\Gn_j = \left[ \,
\begin{matrix}
   \Tn_{j,1} &     \0  		& \0 		& \0       	\\ 
   \0 		  & \Tn_{j,2} 	& \0 		& \0		\\
   \0 		  & \0 			& \Tn_{j,3}& \0		\\
   \0 		  & \0 			& \0 		& \Tn_{j,4}
   \\[1mm]  \cdashline{1-4}[3pt/2pt]  \\[-6.5mm] 
  \Hn_{j,1}^{(2,1)} \Vn_1^{(2,1)} & \Hn_{j,2}^{(2,1)} \Vn_{2}^{(2,1)} & \Hn_{j,3}^{(2,1)} \Vn_{3}^{(2,1)} & \0 \\
  \vdots	 &	\vdots		&	\vdots	& 	\vdots	\\
  \0 & \Hn_{j,2}^{(2,4)} \Vn_2^{(2,4)} & \Hn_{j,3}^{(2,4)} \Vn_{3}^{(2,4)} & \Hn_{j,4}^{(2,4)} \Vn_{4}^{(2,4)} \\
\end{matrix} \, \right].
\label{eq:scB:G43}
\end{IEEEeqnarray}

Thanks to conditions in (\ref{eq:scB:RIA_cond}), all the interference captured during the RIA phase can be removed using the overheard interference from the IIS phase. Now, recall that $\alpha$ represents the number of groups of the RIA phase to which each user belongs. Therefore, the RIA phase provides $\alpha \cdot\Min{N S_2,\varphi} = \alpha \cdot N(M-N)$ extra observations of the desired symbols. Finally, by combining the $NS_1 = \alpha N^2$ linear combinations retrieved from the IIS phase with that obtained during this phase, each receiver obtains $b = \alpha \cdot MN$ LCs of its desired symbols.
\\[2mm]
\textit{Remark}: It can be seen that when $\rho < \roB(G)$ only a subspace of dimension $N(M-N) < \varphi$ of $\Tc_i^{(2,r)}$ is revealed to each receiver. This is in contrast with the case $\rho > \roB(G)$ where the entire subspaces $\Tc_i^{(2,r)}$ must be delivered to $\Rx{i}$ in order to obtaining a sufficient number of observations, and thus ensure linear decodability.

 \subsection{System parameters optimization}
  \label{sec:TGopt}
  
  Given a value of $\rho$, the optimal value of $G$ for the TG scheme may be obtained by means of the steps described in Algorithm 2. The philosophy here is similar to the one in Algorithm 1, and thus its description omitted to avoid redundancy. The parameters, e.g. number of symbols $b$ and number of slots per round $S_1$, $S_2$, given $G$, $K$, and $\rho$, are derived by means of the following DoF optimization problem:
 
\begin{table}[]
\centering
\parbox{.45\linewidth}{
\begin{tabular}{l l}
\toprule
 \multicolumn{2}{c}{{\bf Algorithm 2}: $G$ solver} \\
 \hline  \bigstrut
{\bf Step 1}: & For a given value of $\rho$, find $x\in \{2,\dots,K\}$ minimizing \\
 & ${\rho - \frac{1+\alpha(x) \cdot(x-1)}{1+\alpha(x) \cdot(x-2)}}$, with ${\rho \geq \frac{1+\alpha(x) \cdot(x-1)}{1+\alpha(x) \cdot(x-2)}}$, $\alpha(x) = \binom{K-1}{x-1}$\\[2mm]
{\bf Step 2}: & $ y := \frac{1+\alpha(x) \cdot(x-1)}{K+\pare{x(x-2)+1)}\binom{K}{x}}, \, z := \frac{x+1}{K}\frac{\rho}{\rho + x}$ \\[4mm]
{\bf Step 3}: & $G(\rho) = \left\{\begin{matrix}
      x & x \in \{2,K\} \\
      x & 2 < x < K, \, y > z  \\
      x-1 & \text{otherwise}
     \end{matrix} \right. $\\[7mm]
\hline
\end{tabular}
}
\end{table}

\begin{subequations}
\begin{IEEEeqnarray}{c c l}
 \mathcal{P}_2: & \underset{ \left\{b,S_1,S_2 \right\} \in \enters}{\text{maximize}} \quad  & \frac{1}{N} \frac{b}{K S_1+ \binom{K}{G} S_2} \label{eq:TG:objfunc} \\[2mm]
& s.t. & M S_1 \geq b\label{eq:TG:maxProblem1} \\
  &   & N S_1 < b \label{eq:TG:maxProblem2} \\
    & & NS_2 \leq (G-1)NS_1 - (G-2)b \label{eq:TG:maxProblem3} \\
    & & N\parep{S_1+\alpha(G) \cdot S_2} \geq b \label{eq:TG:maxProblem4}.
\end{IEEEeqnarray}
\label{eq:TG:maxProblem}
\end{subequations} 
While the objective function corresponds to number of symbols delivered per user divided by the duration of the communication, and normalized, the different constraints imposed to ensure linear feasibility are next described:

\vskip 2mm

\subsubsection{Transmit rank during the IIS phase (\ref{eq:TG:maxProblem1})} 
During the first phase, $MS_1$ linear combinations of the $b$ symbols are transmitted using $M$ antennas, and during $S_1$ slots. Then, for linear decodability of the desired symbols, no more symbols than the number of transmit dimensions can be sent, thus we force $ M S_1 \geq b$.

\vskip 2mm

\subsubsection{Need of RIA phase (\ref{eq:TG:maxProblem2})}
Since the first phase provides $NS_1$ interference-free linear observations of the desired symbols, we force $NS_1 < b$.

\vskip 2mm

\subsubsection{Non-redundant RIA phase (\ref{eq:TG:maxProblem3})}
The precoding matrices for each round of the second phase lie on a subspace of dimension $\varphi$, see (\ref{eq:scB:Tdef}) and (\ref{eq:scB:dimIntersection}), and they are used during $S_2$ slots. Then, to avoid redundancy on the received signals, we force that no more than $\varphi$ linear combinations are obtained at the receivers, i.e. $S_2 < \varphi$.
\vskip 2mm

\subsubsection{Linear combinations at the end of the transmission (\ref{eq:TG:maxProblem4})}
Each round of the first phase provides $NS_1$ LCs of desired symbols to each receiver, while each round of the second phase $\Min{ MS_2,NS_2,\varphi} = NS_2$, which follows from $M>N$ the previous constraint. Hence, since each user is active during $\alpha(G)$ rounds of the RIA phase the number of interference-free linear combinations of desired symbols obtained at the end of the transmission are ${NS_1 + \alpha(G) \cdot NS_2}$, and they should be enough for linearly decoding the $b$ desired symbols.
\vskip 2mm
 

This problem will be handled as problem $\mathcal{P}_1$. First, $S_2$ is removed by setting it to its minimum feasible integer value, i.e. 
\begin{IEEEeqnarray}{c}
 S_2^*= \left\lceil{\frac{1}{\alpha(G) } \pare{\frac{b}{N} - S_1}}\right\rceil,
 \label{eq:S2TG}
\end{IEEEeqnarray}
dictated by (\ref{eq:TG:maxProblem4}). Then, (\ref{eq:TG:maxProblem3}) foces that:
\begin{IEEEeqnarray}{c}
 (G-1)NS_1 - (G-2)b \geq N \left\lceil{\frac{1}{\alpha(G) } \pare{\frac{b}{N} - S_1}}\right\rceil \geq N \frac{1}{\alpha(G) } \pare{\frac{b}{N} - S_1},
\end{IEEEeqnarray}
Therefore, $S_1$ may be written as follows:
\begin{IEEEeqnarray}{c} 
S_1^* \! =  \left\lceil b \cdot \max \pare{ {\frac{1}{M}}, \frac{1}{N}\frac{1+\alpha(G) \cdot (G-2)}{1 + \alpha(G) \cdot (G-1)} } \right\rceil,
 \label{eq:S1TG}
\end{IEEEeqnarray}
where $\BI$ and $\BII$ follow from choosing one of the two values above, with the threshold given by
$\roB(G) = \frac{1+\alpha(G) \cdot\pare{G-1}}{1+\alpha(G) \cdot\pare{G-2}}$.

\section{3-user PSR scheme ($M \approx N$)}
\label{sec:PSR} 

The scheme of $P=3$ phases proposed in \cite{Abdoli_IC} for the 3-user SISO IC is generalized to the 3-user MIMO case, proving \mbox{Theorem \ref{th:innerbound3}}. Moreover, \mbox{Theorem \ref{th:innerbound}} for $\left( \rox , \roy(K) \right]$ follows from applying this scheme together with time-sharing concepts. Next section gives an intuition behind this strategy. Then, each of the phases is built, and finally we present the optimization problem that provides the optimal system parameters for any antenna setting and number of users. 
 
\subsection{Overview of the precoding strategy} 
\label{sec:overview}
 
This approach is designed according to the three ingredients exploited so far: {\it delayed CSIT precoding, user scheduling}, and {\it redundancy transmission}. For this reason, it is denoted as the Precoding, Scheduling, Redundancy scheme.
The first and third phases will be labeled as the JIS and RIA phases, as those phases for the RIA scheme. In a similar manner, the objective is to jointly sense the interference for the former and to transmit signals do not causing additional interference for the latter, i.e. aligned with the overheard interference. This is achieved by exploiting only {\it delayed CSIT precoding} and {\it redundancy transmission}, thus all users are active during those phases. But, a {\it hybrid phase} developed by pairs is introduced as the second phase. 
The objective of the hybrid phase is twofold. First, each transmitter based on channel feedback reconstructs the overheard information created at each receiver during the JIS phase to deliver desired linear combinations of symbols. Second, some redundancy is sent in order to create the overheard interference terms that will be used during the last phase. Hence, all the three ingredients are mixed up in this phase in pursuit of DoF maximization. According to all these ideas, we have:
\begin{IEEEeqnarray}{c c c}
R_1 = R_3 = 1, & \quad & R_2 = \binom{3}{2}=3, \\[2mm]
 G_1 = G_3 = 3, & \quad & 
 G_2 = 2,
\end{IEEEeqnarray}
which is also summarized in \mbox{Fig. \ref{fig:frameAbdoli}}. The optimal system parameters are derived in \mbox{Section \ref{sec:BSRopt}}, and specified in \mbox{Table \ref{tab:summary_BSR}}. Recall that $\roBSR{1} \approx 0.7545$,  $\roBSR{2}\approx 0.7847$, see (\ref{eq:roBSR1}) and (\ref{eq:roBSR2}), which means that regimes $\CII$ and $\CIII$ require $M,N>10$. Moreover, it can be seen that this scheme is always outperformed by the RIA scheme for regime $\CI$. Therefore, the most significant finding in this case is that the DoF inner bound for SISO ($\djin = \frac{12}{31}$) is valid whenever $\rho\geq \frac{4}{5}$. Consequently, next sections focus on regime $\CIV$ for simplicity on the description.

\begin{table}[h]
\caption{System parameters for the PSR scheme as a function of $\rho$}
\begin{center}
\begin{tabular}{c | c | c | c | c | c}
  &  $b$  & $S_1$ & $S_2$ & $S_{3}$ & $\djin$  \\[1mm]
\cline{1-6} \bigstrut 
$\CI=\{\frac{1}{2} < \rho \leq \roBSR{1} \} $ &
$M^3$ & $M^2$ & $M(N-M)$ & $2(N-M)^2$ & $\frac{\rho^3}{2-\rho}$ \\
\cline{1-6} \bigstrut 
$\CII=\{\roBSR{1} < \rho \leq \roBSR{2} \} $ &
$2M^2N$ & $2MN$ & $2N(N-M)$ & $5M^2-6MN+2N^2$ & $\frac{2\rho^2}{5\rho^2-10\rho+8}$ \\
\cline{1-6} \bigstrut
$\CIII=\{\roBSR{2} < \rho < \frac{4}{5}\} $ & 
$6MN$ & $6N$ & $4N-3M$ & $4(3M-2N)$ & $\frac{6\rho}{3\rho+10}$ \\
\cline{1-6} \bigstrut
 $\CIV=\{\rho \geq \frac{4}{5} \} $ & $12N$ & 15 & 4 & 4 & $\frac{12}{31} $ \\
\end{tabular} 
\end{center}
\label{tab:summary_BSR}
\end{table} 

\afterpage{
\begin{figure}[]
\begin{minipage}[]{1\linewidth}
  \centering  
  \centerline{\includegraphics[width=0.4\linewidth]{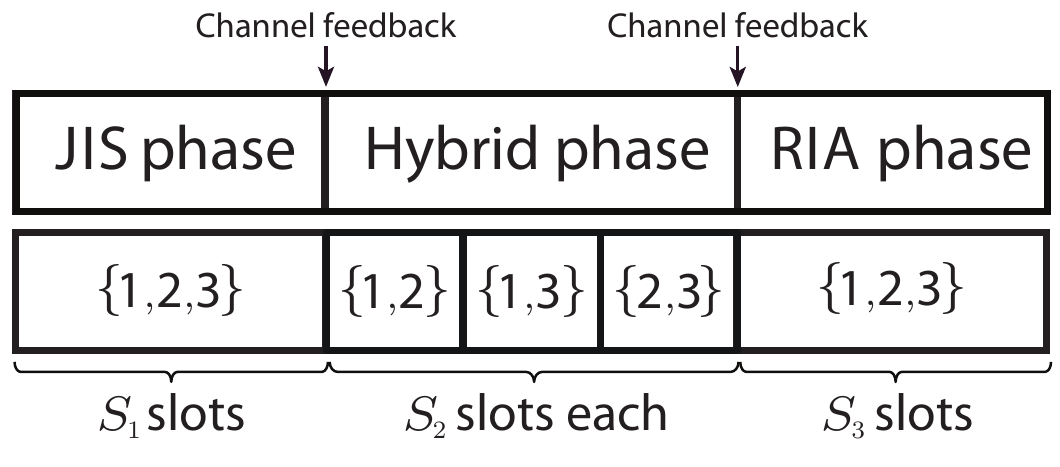}}
\end{minipage}  
\caption{Transmission frame for the PSR scheme. Each phase $p$ has $\binom{K}{p}$ rounds of $S_p$ slots. Active groups $\Ac^{(p,r)}$ are represented for each round of the three phases.} 
\label{fig:frameAbdoli} 
\end{figure} 
}
%

\subsection{Joint interfering sensing phase}
The first phase lasts for $S_1$ slots where transmitters have no CSI, thus they transmit with generic full-rank precoding matrices $\Vn_i^{(1)} \Cmat{MS_1}{b}$ selected from a predetermined dictionary known by all nodes. The development of this phase is exactly the same as for the RIA scheme, with the dimensions specified in \mbox{Table \ref{tab:summary_BSR}}, where $b=12N$, $S_1 = 15$. Then, 
similarly we define the receiving filter ${\Un_{j}^{(1)} \Cmat{2\varphi_1}{NS_1}}$, ${\forall i \neq j}$, with
\begin{IEEEeqnarray}{c} 
{\varphi_1 =  \Min{NS_1-b,b} = 3N}, \label{eq:varphi1}
\end{IEEEeqnarray}
which consists of the composition of two linear filters ${\Un_{j,i}^{(1)} \Cmat{\varphi_1}{NS_1}}, i\neq j$, defined such that (\ref{eq:scA:PIC_ISph}) is satisfied. 
Then, $ {\Tn_{j,i} = \Un_{j,i}^{(1)} \Hn_{j,i}^{\left(1\right)} \Vn_{i}^{\left(1\right)} \Cmat{\varphi_1}{b}},i\neq j$ is again defined representing the residual interference from $\BS{i}$ after applying the linear filter $\Un_{j,i}^{(1)}$, and subspaces 
$ {\Tc_{j,i} = \RSpan{\Tn_{j,i}}}$.

\subsection{Hybrid phase}
The transmission is developed by pairs, where each pair transmits during $S_2 =4$ slots. The objective of this phase is twofold. First, each transmitter exploits the overheard information available at each receiver after the JIS phase to deliver desired linear combinations of symbols, similarly to the second phase of the TG scheme \mbox{(Section \ref{sec:innerboundTG:2})} when $G=2$. Second, each transmitter sends some redundancy in order to create overheard interference that will be seized during the last phase. 

Consider the $(2,r)$th round, with active users $\Ac^{(2,r)} = \{ i,j \}$. The transmitted signals are designed such that
\begin{IEEEeqnarray}{c}
\begin{matrix}
\RSpan{ 
      \Hns{i,j}{2,r} \Vns{j}{2,r} } 
      \subseteq 
      \Tc_{i,j}, \label{eq:BSR:RIA_phase2} \quad
      \RSpan{ 
      \Hns{j,i}{2,r} \Vnsg{i}{2,r} } 
      \subseteq 
      \Tc_{j,i}, 
\end{matrix}
\end{IEEEeqnarray}
thus the precoding matrices are set to
\begin{IEEEeqnarray}{c}
\begin{matrix}
 \Vns{i}{2,r}  =  \boldsymbol{\Sigma}_i^{(2,r)} \Tn_{j,i} \label{eq:BSR:desPhase2}, \quad
 \Vns{j}{2,r}  =  \boldsymbol{\Sigma}_j^{(2,r)} \Tn_{i,j},
\end{matrix}
 \label{eq:BSR:precPhase2}
\end{IEEEeqnarray}
where $\boldsymbol{\Sigma}_i^{(2,r)},\boldsymbol{\Sigma}_j^{(2,r)} \Cmat{MS_2}{\varphi_1}$ are some arbitrary full rank matrices ensuring the transmit power constraint. 
For each active pair, $NS_2 = 4N$ LCs of symbols are received, although the rank of the transmitted signals is 
\begin{IEEEeqnarray}{c}
 {\Rank{\Vns{i}{2,r}}=\Min{MS_2,\Dimp{\Tc_{j,i}}}  = \varphi_1 = 3N},
 \label{eq:BSR:dimPhase2}
\end{IEEEeqnarray}
thus there exists some redundancy on the received signals. In this case the per-phase receiving filters are defined as follows:
\begin{IEEEeqnarray*}{c}
 \Uns{1}{2} = \bdiag{\I,\I,\stack{\Uns{1,2}{2},\Uns{1,3}{2}}}, \\
 \Uns{2}{2} = \bdiag{\I,\stack{\Uns{2,1}{2},\Uns{2,3}{2}},\I}, \\
 \Uns{3}{2} = \bdiag{\stack{\Uns{3,1}{2},\Uns{3,2}{2}},\I,\I}.
\end{IEEEeqnarray*}
where ${\Uns{j,i}{2} \Cmat{\varphi_2}{NS_2}}$, with
\begin{IEEEeqnarray}{c}
{\varphi_2 =  \Min{NS_2-\varphi_1,\varphi_1} = N}. \label{eq:varphi2}
\end{IEEEeqnarray}
Note that the received signal is modified only for the round where all transmitted signals are interference. The objective of this processing is to obtain signal spaces where the desired signals is interefered by only one user, which will be useful to align the interference during the last phase. For example, the processed signal at the first receiver for the $(2,3)$th round writes as:
\begin{IEEEeqnarray}{c}
\zn_1^{(2,3)} = 
\begin{bmatrix}
 \Uns{1,2}{2} \\ \Uns{1,3}{2}
\end{bmatrix}
 \left[
      \Hns{1,2}{2,3}   \Vns{2}{2,3}  , 
      \Hns{1,3}{2,3}   \Vns{3}{2,3}  
 \right] 
 \begin{bmatrix}
  \xn_2 \\ \xn_3
 \end{bmatrix}
 =  
\begin{bmatrix}
   \Fng{3,2}{1} & \0 \\ 
  \0 & \Fng{2,3}{1}
 \end{bmatrix} 
 \begin{bmatrix}
  \xn_2 \\ \xn_3
 \end{bmatrix},
 \label{eq:BSR:PIC_hybridPhase} 
\end{IEEEeqnarray}
where $\Fng{k,i}{j} \Cmat{\varphi_2}{b}$ is defined as
\begin{IEEEeqnarray}{c}
 \Fng{k,i}{j} = \Uns{j,i}{2} \Hns{j,i}{2,r} \Vns{i}{2,r} =  
 \Uns{j,i}{2} \Hns{j,i}{2,r} \boldsymbol{\Sigma}_i^{(2,r)} \Tn_{k,i} \label{eq:BSR:Tfase2}
 \\
 \Fcg{k,i}{j} = \RSpanp{\Fng{k,i}{j}} \subset \Tc_{k,i},
 \label{eq:BSR:incTfase2fase1}
\end{IEEEeqnarray}
i.e. $\Fcg{k,i}{j}$ is the remaining contribution of $\Vns{i}{2,r}$ at $\UE{j}$ after suppressing the signal corresponding to user $k$, i.e. the other active transmitter during the $(2,r)$th round. Moreover, it represents the subspace of $\Tc_{k,i}$ (completely known at $\UE{k}$) that is known thanks to this phase at $\UE{j}$. 
For a better reader's understanding, let us write the signal space matrix obtained at $\UE{1}$ after this phase:
\begin{IEEEeqnarray}{c}
\def\arraystretch{1.8}
\Gn_1^{(2)} = \left[ \,
\begin{matrix}
  \Un_{1,2}^{(1)} \Hn_{1,1}^{\left(1\right)}   \Vn_{1}^{\left(1\right)} & \Tn_{1,2}&     \0          \\ 
   \Un_{1,3}^{(1)} \Hn_{1,1}^{\left(1\right)}   \Vn_{1}^{\left(1\right)} &      \0     & \Tn_{1,3}    
   \\[1mm]  \cdashline{1-3}[3pt/2pt]  \\[-6.5mm] 
  \Hns{1,1}{2,1} \Vns{1}{2,1} & \Hns{1,2}{2,1} \Vns{2}{2,1} & \0 \\
   \Hns{1,1}{2,2} \Vns{1}{2,2} & \0 &  \Hns{1,3}{2,2} \Vns{3}{2,2}\\
   \0  &  \Fng{3,2}{1} & \0 \\
   \0  &      \0       & \Fng{2,3}{1}\\
\end{matrix} \, \right].
\label{eq:BSR:matrixG2}
\end{IEEEeqnarray}
where the dotted lines separate the signals corresponding to each phase, and $\Gn_j^{(p)}$ collects the rows of the signal space matrix $\Gn_j$ up to phase $p$.
 
Finally, the number of interference-free LC of desired signals each receiver can retrieve after this phase is summarized. On the one hand, since at each receiver the signals of each round occupy $NS_2 = 4N$ dimensions, and the interference has rank $\varphi_1=3N$ only, there exists almost surely a $\varphi_2$-dimensional subspace where interference can be projected to. Then, from the two pairs $2 \cdot \Min{\varphi_1, \varphi_2} = 2N$ LCs are obtained. On the other hand, since precoding matrices are designed to align the interference (conditions in (\ref{eq:BSR:RIA_phase2})), $\UE{j}$ will be able to combine the first phase processed signals with the second phase received signals to cancel the interference. Consequently, $2\varphi_1 = 6N$ additional interference-free LCs of desired signals are retrieved, and only $b-8N = 4N$ more LCs are required for ensuring linear decodability.

\subsection{RIA phase}
The third phase lasts for $S_3 = 4$ slots, where all users are active. The objective is to design the transmitted signals based on the information commonly known at the non-intended receivers after the first two phases. The precoding matrices for this phase are constructed as follows:
\begin{IEEEeqnarray}{c c c c c}
 \Vn_{1}^{(3)}  =  \boldsymbol{\Sigma}_1^{(3)}
 \begin{bmatrix}\Fng{2,1}{3} \\ \Fng{3,1}{2} \end{bmatrix} \label{eq:BSR:precRIA}, &
 \quad &
 \Vn_{2}^{(3)}  =  \boldsymbol{\Sigma}_2^{(3)}
 \begin{bmatrix}\Fng{1,2}{3} \\ \Fng{3,2}{1} \end{bmatrix}  &
 \quad & 
 \Vn_{3}^{(3)}  =  \boldsymbol{\Sigma}_3^{(3)}
 \begin{bmatrix}\Fng{1,3}{2} \\ \Fng{2,3}{1} \end{bmatrix}
\end{IEEEeqnarray}
where $\boldsymbol{\Sigma}_i^{(3)} \Cmat{MS_3}{2\varphi_2}$. This design ensures that all the generated interference is already known at both non-intended receivers, thus receivers will be able to remove it.
Moreover, each receiver observes $NS_3 = 4N$ linear combinations of the transmitted signals of rank
\begin{IEEEeqnarray}{r l}
\Rank{\Vn_i^{(3)}} = & \, \,\Dim{\Fcg{j,i}{k}+\Fcg{k,i}{j}} \\
   = & \, \,\Dim{\Fcg{j,i}{k}}+\Dim{\Fcg{k,i}{j}} =2\varphi_2 = 2N, \, i \neq j \neq k. 
   \label{eq:BSR:precPhase3}
\end{IEEEeqnarray}
Then, the same idea as for the second phase applies here: some redundancy is transmitted in order to apply zero-forcing concepts at the receiver. Following the same notation as before, two linear filters $\Uns{j,i}{3} \Cmat{\varphi_3}{NS_3}, j\neq i,$ are applied at each receiver, with 
\begin{IEEEeqnarray}{c}
{\varphi_3 =  \Min{NS_3-2\varphi_2,2\varphi_2} = 2N}, \label{eq:varphi3}
\end{IEEEeqnarray}
For brevity and clarity, the final signal space matrix at $\UE{1}$ is next shown:
\begin{IEEEeqnarray}{c}
\def\arraystretch{1.8}  
\Gn_1 = \left[ \,
\begin{matrix}
   \Un_{1,2}^{(1)} \Hn_{1,1}^{\left(1\right)}   \Vn_{1}^{\left(1\right)} & \Tn_{1,2}&     \0          \\ 
   \Un_{1,3}^{(1)} \Hn_{1,1}^{\left(1\right)}   \Vn_{2}^{\left(1\right)} &      \0     & \Tn_{1,3}     
   \\[1mm]  \cdashline{1-3}[3pt/2pt]  \\[-6.5mm] 
  \Hns{1,1}{2,1} \Vns{1}{2,1} & \Hns{1,2}{2,1} \Vns{2}{2,1} & \0 \\
   \Hns{1,1}{2,2} \Vns{1}{2,2} & \0 &  \Hns{1,3}{2,2} \Vns{3}{2,2}\\
   \0  &  \Fng{3,2}{1}  & \0 \\
   \0  &  \0 & \Fng{2,3}{1} 
   \\[1mm]  \cdashline{1-3}[3pt/2pt]  \\[-6.5mm] 
   \Uns{1,2}{3}\Hns{1,1}{3} \Vns{1}{3} & \Uns{1,2}{3}\Hns{1,2}{3} \Vns{2}{3} & \0 \\
   \Uns{1,3}{3} \Hns{1,1}{3} \Vns{1}{3} & \0 &  \Uns{1,3}{3} \Hns{1,3}{3} \Vns{3}{3}
\end{matrix} \, \right],
\label{eq:BSR:matrixG3}
\end{IEEEeqnarray}
where the signals received during the RIA phase are processed using $\Uns{1,2}{3}$ and $\Uns{1,3}{3}$, see the last two blocks rows. Now it is easy to see that all the interference is aligned. For example, consider the 1st, 5th and 7th block rows. Since 
\begin{IEEEeqnarray}{c}
 \RSpan{\Uns{1,2}{3}\Hns{1,2}{3} \Vns{2}{3}} \subseteq \RSpan{\Vns{2}{3}}, \label{eq:BSR:RIA_phase3}\\
 \RSpan{\Vns{2}{3}} \subseteq \Fcg{3,2}{1} + \Fcg{1,2}{3}, \\
 \Fcg{1,2}{3} \subset \Tc_{1,2} \label{eq:BSR:T2cT1},
\end{IEEEeqnarray}
the signals corresponding to the 1st and 5th block rows can be used to remove the interference from the signals represented by the 7th block row. Then, $2\varphi_2$ LCs of desired signals are retrieved. Following similar arguments for rows 2nd, 6th, and 8th, $2N$ extra LCs are obtained. Combining the $4\varphi_2 = 4N$ LCs of desired signals obtained from this phase with the $8N$ LCs from previous phases, each receiver obtains enough LCs for linearly decode all of its $b=12N$ desired symbols.

 \subsection{System parameters optimization}
  \label{sec:BSRopt}
  
  The parameters for the PSR scheme are derived by means of the following DoF optimization problem:
  \begin{subequations} \label{eq:BSR:maxProblem}
\begin{IEEEeqnarray}{c c l}
  \mathcal{P}_3: & \underset{ \left\{b,\varphi_1,\varphi_2, \varphi_3\right\} >0}{\text{maximize}} \quad  & \frac{b}{b+4\varphi_1+5\varphi_2+\varphi_3} \label{eq:BSR:objfunc} \\[2mm]
& s.t. & \rho\pare{\varphi_1+b} \geq b \label{eq:BSR:maxProblem1} \\
  &   & 4\varphi_1 \geq b \label{eq:BSR:maxProblem2} \\
  &   & \rho\pare{\varphi_1+\varphi_2} \geq \varphi_1 \label{eq:BSR:maxProblem3} \\
   &  & \varphi_2 \leq \varphi_1 \label{eq:BSR:maxProblem4} \\
   &  & 2\pare{\varphi_1+\varphi_2}<b  \label{eq:BSR:maxProblem5} \\
   &  & \rho \pare{\varphi_3 +2\varphi_2} \geq 2\varphi_2 \label{eq:BSR:maxProblem6} \\
   &  & 2\pare{\varphi_1 + \varphi_2 + \varphi_3} \geq b \label{eq:BSR:maxProblem7} \\
   &  & \varphi_3 \leq 2 \varphi_2 \label{eq:BSR:maxProblem8}  
\end{IEEEeqnarray}
\end{subequations}
formulated in terms of $\varphi_i>0, i=1,2,3$, where the number of slots can be retrieved by applying the following change of variables:
\begin{IEEEeqnarray}{c}
 \varphi_1 = NS_1 - b, \quad \varphi_2 = NS_2 - \varphi_1, \quad \varphi_3 = NS_3 - 2\varphi_2.
\end{IEEEeqnarray}
While the objective function corresponds to $\frac{b}{N\tau}$ in terms of the new variables, the constraints imposed to ensure linear feasibility are next described:

\vskip 2mm

\subsubsection{Transmit rank during the JIS phase (\ref{eq:BSR:maxProblem1})} 
Similarly to other schemes, $ M S_1 \geq b$ is imposed to ensure the transmit rank.

\vskip 2mm

\subsubsection{Linear combinations on the system (\ref{eq:BSR:maxProblem2})}
After the first phase processing, $4\varphi_1$ linear combinations of the symbols of each user are distributed along the receivers: $2\varphi_1$ at the intended receiver (known coupled with interference), and $\varphi_1$ at each non-intended receiver. Then, since the rest of phases are just retransmissions, a necessary condition is that at least obtaining all of them the $b$ desired symbols should be linearly decodable.

\vskip 2mm

\subsubsection{Transmit rank during the hybrid phase (\ref{eq:BSR:maxProblem3})}
Written in terms of the new variables, it is forced $MS_2 \geq \varphi_1$, since the rank of the transmitted signals during each second phase round is equal to $\varphi_1$, see (\ref{eq:BSR:dimPhase2}).
\vskip 2mm

\subsubsection{Bounded redundancy during the hybrid phase and need of RIA phase (\ref{eq:BSR:maxProblem4}) and (\ref{eq:BSR:maxProblem5})}
After the hybrid phase, each receiver is able to retrieve $\varphi_1 + \varphi_2$ interference-free LCs of desired symbols from each of the two rounds where desired LCs of signals are sent. First, exploiting the redundancy on the received signals due to $\varphi_2 = NS_2 - \varphi_1 >0$, $\Min{\varphi2,\varphi1}$ linear combinations can be retrieved by zero-forcing concepts. Then, we force  (\ref{eq:BSR:maxProblem4}), since having $\varphi_2 > \varphi_1$ does not provide additional LCs. This constraint bounds the value of $S_2$, and it is also imposed by $\Fcg{k,i}{j} \subset \Tc_{k,i}$, as assumed in (\ref{eq:BSR:incTfase2fase1}).

On the other hand, $\varphi_1$ LCs are obtained through RIA concepts, by projecting the signals of the corresponding round of the hybrid phase onto a subspace of dimension $\varphi_1$, and combining them with the JIS phase processed signals. Consequently, at the end of the hybrid phase $2(\varphi_1 + \varphi_2)$ independent observations are obtained. (\ref{eq:BSR:maxProblem5}) ensures that still some extra LCs are required, and thus RIA phase is necessary.
\vskip 2mm

\subsubsection{Transmit rank during the RIA phase (\ref{eq:BSR:maxProblem6})}
Written in terms of the new variables, it is forced $MS_3 \geq 2\varphi_2$, see (\ref{eq:BSR:precPhase2}).
\vskip 2mm

\subsubsection{Linear combinations at the end of the transmission (\ref{eq:BSR:maxProblem7}) and bounded redundancy during the RIA phase (\ref{eq:BSR:maxProblem8}) }
The signal received during the RIA phase is processed to decouple the interference, see (\ref{eq:BSR:matrixG3}). Those processed signals combined with the rest of available overheard interference provide $2\cdot\Min{\varphi_3,2\varphi_2}$ extra observations of the desired symbols. First, (\ref{eq:BSR:maxProblem8}) is forced to bound the value of $S_3$, and because in this case more redundancy does not provide additional LCs. Second, the number of interference-free LCs of desired signals each receiver is able to retrieve at the end of the transmission is equal to $2 \cdot \pare{\varphi_1+\varphi_2+\varphi_3}$, and it should be enough to linearly decode all the $b$ desired symbols.
\vskip 2mm

The problem $\mathcal{P}_3$ in (\ref{eq:BSR:maxProblem}) is next solved. 
Before proceeding, let us introduce the following proposition:
\vskip 4mm

\begin{proposition}
\label{prop:FME}
 Consider the following two linear inequalities:
 \begin{IEEEeqnarray}{c}
    x + b  y \geq c  z \label{eq:propa}, \\
   d x + e y \leq f  z \label{eq:propb},
 \end{IEEEeqnarray}
where $\{a,b,c,d,e,f\}$ are positive given parameters, and $\{x,y,z\}$ represent unknown variables. Then, any solution satisfying both inequalities also satisfies:
 \begin{IEEEeqnarray}{c}
   c  d  x + c e  y \leq  f a x + f b  y. \label{eq:propc}
 \end{IEEEeqnarray}
\end{proposition}

\vskip 4mm

This trivial proposition is useful because it allows suppressing variables from linear constraints. 
Actually, it is the basis of the Fourier-Motzkin Elimination method, see \cite{FMelim}.

Consider the application of Proposition 1 to (\ref{eq:BSR:maxProblem6}), (\ref{eq:BSR:maxProblem7}), and (\ref{eq:BSR:maxProblem8}), such that variable $\varphi_3$ is removed. This leads to the following two constraints:
\begin{IEEEeqnarray}{c}
2\pare{\varphi_1+3\varphi_2} \geq b, \label{eq:BSR:maxProblem7p}\\
\varphi_2 \pare{1-2\rho} \geq 0,
\end{IEEEeqnarray}
where the second constraint forces $\rho \geq \frac{1}{2}$.
Now, let us apply again the proposition to (\ref{eq:BSR:maxProblem4}), (\ref{eq:BSR:maxProblem5}), and the new constraint (\ref{eq:BSR:maxProblem7p}) in order to remove $\varphi_2$. Again, two new constraints are procuded:
\begin{IEEEeqnarray*}{c} 
8 \varphi_1 \geq b \\
\varphi_1 \pare{1-2\rho} \geq 0,
\end{IEEEeqnarray*}
which are loose with respect to the rest of constraints. Then, the value of $\varphi_1$ is completely determined by (\ref{eq:BSR:maxProblem1}) and (\ref{eq:BSR:maxProblem2}), as follows:
\begin{IEEEeqnarray}{c}
	\varphi_1 = b\Max{\frac14,\frac{1-\rho}{\rho}},
	\label{eq:BSR:optimalfi1}
\end{IEEEeqnarray}
thus establishing two regions: $\rho \geq \frac{4}{5}$ and $\rho < \frac{4}{5}$. For a given value of 
$\varphi_1$, the optimal $\varphi_2$ is decided according to (\ref{eq:BSR:maxProblem4}) and (\ref{eq:BSR:maxProblem7p}), as follows:
\begin{IEEEeqnarray}{c}
	\varphi_2 =  \Max{\varphi_1 \frac{1-\rho}{\rho}, \frac16 \pare{b-2\varphi_1}}.
	\label{eq:BSR:optimalfi2}
\end{IEEEeqnarray}
Finally, the optimal value of $\varphi_3$ is set according to
\begin{IEEEeqnarray}{c}
\varphi_3 =  \Max{2 \varphi_2 \frac{1-\rho}{\rho}, \frac{b}{2} - \varphi_1 -\varphi_2}.
	\label{eq:BSR:optimalfi3}
\end{IEEEeqnarray}
It can be checked that the control constraints (\ref{eq:BSR:maxProblem5}) and (\ref{eq:BSR:maxProblem6}) are always satisfied following these rules. The values in \mbox{Table \ref{tab:summary_BSR}} are obtained by inverting the change of variables and taking the value of $b$ such that $S_1$, $S_2$, and $S_3$ are integer values.

\section{DoF-delay trade-off}
\label{sec:practical}

The precoding schemes exploiting delayed CSIT require multi-phase transmissions. For some settings, this entails long communication delays, and a high number of transmitted symbols, thus increasing the complexity of the encoding/decoding operation at transmitters/receivers. This section studies the DoF-delay trade-off of the proposed and some state-of-the-art schemes.
Thanks the DoF-delay trade-off analysis, two main insights are concluded:{
\begin{itemize}
 \item The supremacy in terms of achievable DoF of one scheme w.r.t. another depends on the allowed complexity of the tranmission, i.e. number of transmitted symbols or duration of the communication. For example, when $\rho=1$, $K=3$, the RIA scheme outperforms the PSR scheme given a maximum number of time slots allowed for the communication.
 \item The communication delay can be highly alleviated without high DoF penalties. Many examples  are provided showing the balance between optimal (but usually large) parameters and maximum DoF w.r.t. practical parameters and competitive achievable DoF.
\end{itemize}}

Two methodologies are used in the sequel to study the DoF-delay trade-off of the proposed schemes. First, the following three sections analyze this trade-off by limiting the maximum number of symbols per user that can be transmitted to $\Bmax$, i.e. by introducing the following constraint into the system parameters optimization problems:
\begin{IEEEeqnarray*}{c}
 b \leq B.
\end{IEEEeqnarray*}
The case where this constraint is omitted or, equivalently, $\Bmax \to \infty$, will be hereafter denoted as the {\it unbounded case}. 
For each scheme, a simplified version of the DoF optimization problem for finite $\Bmax$ is provided. Then, at least two examples are evaluated for each case, one for $K=3$ and one for $K=6$, which are useful to benchmark one of the values of $\rho$ for \mbox{Fig. \ref{fig:mainResults3} and \ref{fig:mainResults}} as a function of $\Bmax$.

On the other hand, an alternative approach is proposed in order to compare the proposed schemes to the PSR scheme in \cite{Abdoli_IC} and its extension to MISO in \cite{Hao_BSR_MISOIC} for $K>3$, addressed in \refbox{Section}{sec:order}. Notice that the first methodology could also be used, but currently we do not have derived DoF optimization problems for those schemes, which remains as future work.

Adopting this second methodology, the DoF-delay trade-off is studied by limiting the order of the transmitted symbols. Although the formulation up to this section works with order-1 symbols and order-1 DoF, the works in the literature usually follow the high-order symbol framework, see e.g. \cite{MAT}, briefly summarized next. 

An order-$m$ symbol refers to a supersymbol which is desired or available at $m$ receivers, either to remove interference or because it contains the symbols intended to that receiver. For example, during the second phase of the RIA and TG schemes, it can be interpreted that order-$L$ and order-$G$ symbols are transmitted, respectively. In contrast to our schemes, usually $K$ phases are scheduled such that order-$m$ symbols are transmitted during the phase $m$, in turn generating a number of order-$(m+1)$ symbols to be delivered during the next phase. This process ends up at phase $K$ where symbols of order-$K$, i.e. intended to all the users, are transmitted by means of TDMA, and not producing additional high-order symbols. Accordingly, the order-$m$ DoF are defined as the efficiency of transmitting order-$m$ symbols through the network. Notice that they account for the DoF of transmitting order-$m$ symbols without interpretation of which user they are intended to. Hence, during this section we work with the sum-DoF, i.e.  $\din = K\djin$. Therefore,
$d^{(m,\text{in})}$ denotes the order-$m$ sum-DoF inner bound, and according to previous arguments they may be formulated in a recursive way as follows \cite{MAT}:
\begin{IEEEeqnarray}{c}
 d^{(m,\text{in})} = f \left(d_j^{(m+1)} \right), \, m=1 \dots K-1, \\
 d^{(1,\text{in})} \equiv d_j^{(\text{in})}, \\
 d^{(K,\text{in})} = 1,
\end{IEEEeqnarray}
i.e. the efficiency of transmitting order-$m$ symbols depends on the efficiency of transmitting order-($m+1$) symbols. Notice that the sum-DoF of order-$K$ are equal to one since we work with normalized DoF.

Inspired by this formulation, we propose to bound the schemes in \cite{Abdoli_IC} and \cite{Hao_BSR_MISOIC} to maximum order $\Theta$ by forcing:
\begin{IEEEeqnarray}{c}
 d_j^{(\Theta,\text{in})} = 1.
\end{IEEEeqnarray}

\subsection{RIA scheme}
\label{sec:RIAsens}

A closed-form solution for $S_1$ and $S_2$ was obtained in \mbox{Section \ref{sec:RIAopt}}, see (\ref{eq:RIA:optS2}) and (\ref{eq:RIA:optS1}). For unbounded $b$, the value of $L$ was obtained given $\rho$ and $K$ by means of Algorithm 1. However, for finite $\Bmax$ the optimal value of $L$ becomes a function of $\Bmax$. In this regard, the achievable DoF for the RIA scheme write as follows:
\begin{IEEEeqnarray}{c}
	 d_j(\Bmax) = \frac{1}{KN} \Max{f_{\text{RIA},1} (\Bmax),f_{\text{RIA},2} (\Bmax)},\label{eq:fRIA} \\
 f_{\text{RIA},1} (x) = \underset{ b \leq x,L}{\text{maximize}} \frac{bL}{\ceil{\frac{b}{M}} +\ceil{\frac{b}{N}}}, \label{eq:fRIA1}\\
 f_{\text{RIA},2} (x) = \underset{ b \leq x,L}{\text{maximize}} \frac{bL}{\ceil{\frac{b}{N}\frac{L^2-L-1}{L}} +\ceil{\frac{b}{N}}},\label{eq:fRIA2} 
\end{IEEEeqnarray}
where $f_{\text{RIA},1} (x)$ and $f_{\text{RIA},2} (x)$ represent the achievable DoF for each side of the stepping function in Figs. \ref{fig:mainResults3} and \ref{fig:mainResults}, or in Theorem \ref{th:innerbound}. Since the value of $L$ depends on $\Bmax$, it is not possible to derive a threshold as $\roA(L)$. Then, we maximize w.r.t. $L$ and $b$, and then just take the maximum between the two sides of the stepping function. 

The maximization problem for finite $\Bmax$ has been solved for the two settings: $(M,N,K)= (4,7,3)$, and $(M,N,K)= (3,4,6)$, where the solutions follow the expressions given in (\ref{eq:fRIA})-(\ref{eq:fRIA1}). The achievable DoF w.r.t. the communication delay are depicted in \refbox{Fig}{fig:DoF_Bmax}-top for $\Bmax = 1 \dots b^*$, where $b^*$ denotes for each case the optimal value of $b$ for the unbounded case. Moreover, the DoF achieved without the need of CSIT are also included for comparison. First, notice that since $L \in \{3,\dots,K\}$, the only possible value for the first setting is $L=3$. In such a case, since $\rho < \roA(3)=\frac{3}{5}$, it follows 
\begin{IEEEeqnarray*}{c}
 d_j(\Bmax)=\frac{1}{KN} f_{\text{RIA},1} (\Bmax).
\end{IEEEeqnarray*}
The more interesting conclusion from this analysis is that the number of required slots can be dramatically reduced without high DoF penalties. In particular, the number of time slots may be halved (from 11 to 5), while 94\% of the maximum DoF are attained (from 0.3636 to 0.3429).
In contrast, for the setting $(M,N,K)= (2,5,6)$ the value of $L$ changes as a function of $\Bmax$, as highlighted in \refbox{Fig}{fig:DoF_Bmax}, top-right. Notice that in this case the number of slots required to outperform TDMA is huge, and DoF gains are insignificant.
  
\pgfplotsset{
plotoptsPropIn/.style={color=orange, thick,mark=*},
plotoptsPrevIn/.style={color=blue, thick,mark=*, mark options={fill=blue,solid}},
plotoptsUp/.style={color=red, thick,mark=*, mark options={fill=red,solid}},
plotoptsNoCSIT/.style={color=black, thick,mark=*, mark options={fill=white,solid}},
} 


 \begin{figure}[]
 \begin{minipage}{\linewidth}
\begin{minipage}[]{0.48\linewidth}
  \centering 


  \newcommand{\dmin}{0}  \newcommand{\dmax}{0.4}
  \newcommand{\taumin}{0} \newcommand{\taumax}{15}
  \newcommand{\conf}{473} 
\pgfplotstableread[col sep=tab]{DoF\conf.txt} \dof

\begin{tikzpicture} 
\begin{axis}[
  ymin=\dmin,ymax=\dmax,xmin=\taumin,xmax=\taumax,
  xmajorgrids,
   ymajorgrids,
   grid style={dashed, gray!30},
  ylabel style={at={(0.02,0.5)},rotate=-90},
  width=\linewidth,   font=\scriptsize,
  xlabel=Total number of slots ($\tau$),  ylabel=$\djin$,
  legend style={at={(0.95,0.05)},anchor=south east,font=\scriptsize}],
  ]
  
   \addplot[plotoptsPropIn] 
  table[x =Rsum, y = d]   from \dof ; 
  \addlegendentryexpanded{RIA};
  \addplot[plotoptsNoCSIT] 
  table[x =RsumnoCSIT, y = dnoCSIT]   from \dof ;   
  \addlegendentryexpanded{TDMA};
  
  \textbox{3}{0.38}{\scriptsize $(M,N,K)=(4,7,3)$}
  
\end{axis}
\end{tikzpicture} 
\end{minipage}  \hspace{3mm}
\begin{minipage}{0.48\linewidth}
  \centering 


  \newcommand{\dmin}{0.1}  \newcommand{\dmax}{0.2}
  \newcommand{\taumin}{0} \newcommand{\taumax}{175}
  \newcommand{\conf}{346} 
\pgfplotstableread[col sep=tab]{DoF\conf.txt} \dof

\begin{tikzpicture} 
\begin{axis}[
  ymin=\dmin,ymax=\dmax,xmin=\taumin,xmax=\taumax,
  xmajorgrids,
   ymajorgrids,
   grid style={dashed, gray!30},
  ylabel style={at={(0.02,0.5)},rotate=-90},
  width=\linewidth,   font=\scriptsize,
  xlabel=Total number of slots ($\tau$),  ylabel=$\djin$,
  legend style={at={(0.95,0.05)},anchor=south east,font=\scriptsize}],
  ]
  
   \addplot[plotoptsPropIn] 
  table[x =Rsum, y = d]   from \dof ; 
  \addlegendentryexpanded{RIA};
  \addplot[plotoptsNoCSIT]  
  table[x =RsumnoCSIT, y = dnoCSIT]   from \dof ;   
  \addlegendentryexpanded{TDMA};
  
  \draw[rotate=12,dashed](149,61) ellipse (42 and 10);
  \textboxx{130}{0.16}{$L=3$}
  
  \draw[rotate=-10,dashed] (11,40) ellipse (10 and 42);
  \textboxx{47}{0.14}{$L=5$}
  
  \textbox{35}{0.195}{\scriptsize $(M,N,K)=(3,4,6)$}
  
\end{axis}
\end{tikzpicture} 
\end{minipage}  
\end{minipage}
\vskip 1mm
\begin{minipage}{\linewidth}
\begin{minipage}[]{0.48\linewidth}
  \centering 


  \newcommand{\dmin}{0.1}  \newcommand{\dmax}{0.5}
  \newcommand{\taumin}{0} \newcommand{\taumax}{25}
  \newcommand{\conf}{753} 
\pgfplotstableread[col sep=tab]{DoF\conf.txt} \dof

\begin{tikzpicture} 
\begin{axis}[
  ymin=\dmin,ymax=\dmax,xmin=\taumin,xmax=\taumax,
  xmajorgrids,
   ymajorgrids,
   grid style={dashed, gray!30},
  ylabel style={at={(0.02,0.5)},rotate=-90},
  width=\linewidth,   font=\scriptsize,
  xlabel=Total number of slots ($\tau$),  ylabel=$\djin$,
  legend style={at={(0.95,0.05)},anchor=south east,font=\scriptsize}],
  ]
  
   \addplot[plotoptsPropIn] 
  table[x =Rsum, y = d, restrict x to domain = 1:\taumax]   from \dof ; 
  \addlegendentryexpanded{TG};
  \addplot[plotoptsPrevIn] 
  table[x =Rsumv, y = dv, restrict x to domain = 1:\taumax]   from \dof ; 
  \addlegendentryexpanded{\cite{Vaze2IC}};
  \addplot[plotoptsNoCSIT] 
  table[x =RsumnoCSIT, y = dnoCSIT]   from \dof ;   
  \addlegendentryexpanded{TDMA};
  
  \draw[rotate=0,dashed](60,120) ellipse (15 and 34); 
  \textboxx{9.5}{0.2}{$G=3$}
  
  \draw[rotate=0,dashed] (68,255) ellipse (15 and 72);
  \textboxx{10}{0.42}{$G=2$}
  
  \textbox{5}{0.48}{\scriptsize $(M,N,K)=(7,5,3)$}
  
\end{axis}
\end{tikzpicture} 
\end{minipage}  \hspace{3mm}
\begin{minipage}{0.48\linewidth}
  \centering 


  \newcommand{\dmin}{0}  \newcommand{\dmax}{0.3}
  \newcommand{\taumin}{0} \newcommand{\taumax}{80}
  \newcommand{\conf}{416} 
\pgfplotstableread[col sep=tab]{DoF\conf.txt} \dof

\begin{tikzpicture} 
\begin{axis}[
  ymin=\dmin,ymax=\dmax,xmin=\taumin,xmax=\taumax,
  xmajorgrids,
   ymajorgrids,
   grid style={dashed, gray!30},
  ylabel style={at={(0.02,0.5)},rotate=-90},
  width=\linewidth,   font=\scriptsize,
  xlabel=Total number of slots ($\tau$),  ylabel=$\djin$,
  legend style={at={(0.95,0.05)},anchor=south east,font=\scriptsize}],
  ]
  
    \addplot[plotoptsPropIn] 
  table[x =Rsum, y = d, restrict x to domain = 1:\taumax]   from \dof ; 
  \addlegendentryexpanded{TG};
  \addplot[plotoptsPrevIn] 
  table[x =Rsumv, y = dv, restrict x to domain = 1:\taumax]   from \dof ; 
  \addlegendentryexpanded{\cite{Vaze2IC}};
  \addplot[plotoptsNoCSIT] 
  table[x =RsumnoCSIT, y = dnoCSIT]   from \dof ;   
  \addlegendentryexpanded{TDMA};
  
  \textbox{16}{0.285}{\scriptsize $(M,N,K)=(4,1,6)$}
  
\end{axis} 
\end{tikzpicture} 
\end{minipage}  
\end{minipage}
\vskip 1mm
\begin{minipage}{\linewidth}
\begin{minipage}[]{0.48\linewidth}
  \centering 


  \newcommand{\dmin}{0.3}  \newcommand{\dmax}{0.4}
  \newcommand{\taumin}{0} \newcommand{\taumax}{35}
  \newcommand{\conf}{113} 
\pgfplotstableread[col sep=tab]{DoF\conf.txt} \dof

\begin{tikzpicture} 
\begin{axis}[
  ymin=\dmin,ymax=\dmax,xmin=\taumin,xmax=\taumax,
  xmajorgrids,
   ymajorgrids,
   grid style={dashed, gray!30},
  ylabel style={at={(0.02,0.5)},rotate=-90},
  width=\linewidth,   font=\scriptsize,
  xlabel=Total number of slots ($\tau$),  ylabel=$\djin$,
  legend style={at={(0.95,0.05)},anchor=south east,font=\scriptsize}],
  ]
  
   \addplot[plotoptsPropIn] 
  table[x =Rsum, y = d,restrict x to domain = 1:\taumax]   from \dof ; 
  \addlegendentryexpanded{RIA};
  \addplot[plotoptsPrevIn] 
  table[x =RsumBSR, y = dBSR,restrict x to domain = 1:\taumax]   from \dof ; 
  \addlegendentryexpanded{3-user PSR};
  \addplot[plotoptsNoCSIT] 
  table[x =RsumnoCSIT, y = dnoCSIT]   from \dof ;   
  \addlegendentryexpanded{TDMA};
  \textbox{7}{0.395}{\scriptsize $(M,N,K)=(1,1,3)$}
  
\end{axis}
\end{tikzpicture} 
\end{minipage}  \hspace{3mm}
\begin{minipage}{0.48\linewidth}
  \centering 


  \newcommand{\dmin}{0}  \newcommand{\dmax}{0.42}
  \newcommand{\taumin}{0} \newcommand{\taumax}{35}
  \newcommand{\conf}{453} 
\pgfplotstableread[col sep=tab]{DoF\conf.txt} \dof

\begin{tikzpicture} 
\begin{axis}[
  ymin=\dmin,ymax=\dmax,xmin=\taumin,xmax=\taumax,
  xmajorgrids,
   ymajorgrids,
   grid style={dashed, gray!30},
  ylabel style={at={(0.02,0.5)},rotate=-90},
  width=\linewidth,   font=\scriptsize,
  xlabel=Total number of slots ($\tau$),  ylabel=$\djin$,
  legend style={at={(0.95,0.05)},anchor=south east,font=\scriptsize}],
  ]
  
   \addplot[plotoptsPropIn] 
  table[x =Rsum, y = d]   from \dof ; 
  \addlegendentryexpanded{RIA};
  \addplot[plotoptsPrevIn] 
  table[x =RsumBSR, y = dBSR,restrict x to domain = 1:\taumax]   from \dof ; 
  \addlegendentryexpanded{3-user PSR};
  \addplot[plotoptsNoCSIT]  
  table[x =RsumnoCSIT, y = dnoCSIT,restrict x to domain = 1:\taumax]   from \dof ;   
  \addlegendentryexpanded{TDMA};
  
  \textbox{7}{0.402}{\scriptsize $(M,N,K)=(4,5,3)$}
  
\end{axis}
\end{tikzpicture} 
\end{minipage}  
\end{minipage}
\caption{Achievable DoF of the proposed schemes vs duration of the transmission $\tau$ for different values of $\Bmax$. The DoF achieved without the need of CSIT or using previous schemes in the literature are also depicted for comparison purposes.}
\label{fig:DoF_Bmax}
\end{figure}

The reader may have noticed that the cases with $\rho > \frac{3}{5}$ have been omitted. In this regard, two additional examples will be shown for the RIA scheme in \mbox{Section \ref{sec:BSRsens}}, deferred to that section in order to compare together the RIA and PSR schemes performance for limited $\Bmax$.

\subsection{TG scheme}
\label{sec:TGsens}
Closed form solutions for $S_1^*$ and $S_2^*$ were found in \mbox{Section \ref{sec:TGopt}}, see (\ref{eq:S2TG}) and (\ref{eq:S1TG}), next restated for reader's convenience. Note that $S_2^*$ depends on the value taken for $S_1^*$, which depend on the antenna ratio and $\Bmax$. In this case, the achievable DoF for a given $\Bmax$ write as follows:
\begin{IEEEeqnarray}{c} 
  d_j(\Bmax) = \underset{ b \leq \Bmax,G}{\text{maximize}} \,\, \frac{1}{N} \frac{b}{K S_1^*+ \binom{K}{G} S_2^*}, \label{eq:fTG} \\[3mm]
 S_1^* \! =  \left\lceil b \cdot \max \pare{ {\frac{1}{M}}, \frac{1}{N}\frac{1+\alpha(G) \cdot (G-2)}{1 + \alpha(G) \cdot (G-1)} } \right\rceil ,\label{eq:fTG-1}\\[3mm]
 S_2^* = \left\lceil \frac{1}{\alpha(G) } \pare{\frac{b}{N} - S_1^*} \right\rceil
  \label{eq:fTG2}, \,\,\, \alpha(G) = \binom{K-1}{G-1}.
\end{IEEEeqnarray}
Two settings are simulated and shown in \refbox{Fig.}{fig:DoF_Bmax}-middle: $(M,N,K)= (7,5,3)$, and $(M,N,K)= (4,1,6)$. While the curves have been obtained by solving the problem $\mathcal{P}_2$ in (\ref{eq:TG:maxProblem}), one can check that they follow the expressions in (\ref{eq:fTG})-(\ref{eq:fTG2}). For comparison purposes, in addition to the TDMA performance, the scheme in \cite{Vaze2IC} for the 2-user IC has been considered. This scheme is applied to the $K$-user case by means of time-sharing, which dramatically increases the communication delay. In order to obtain its performance for different values of $\Bmax$, a DoF maximization problem has been formulated. The problem is very similar to the TG scheme with $G=2$, and thus omitted. 

Both figures show the DoF gains provided by the wise use of delayed CSIT w.r.t. no CSIT by increasing the duration of the communication $\tau$. Two remarkable observations can be drawn, one for each setting. For the setting $(M,N,K)= (7,5,3)$ the DoF attained using delayed CSIT for both strategies are similar for the unbounded case. However, this is at the cost of a high communication delay for the scheme in \cite{Vaze2IC}. If otherwise $\tau$ is reduced, then the TG scheme clearly outperforms any other strategy.

On the other hand, for the setting $(M,N,K)= (4,1,6)$, it can be observed that the unbounded case requires $\tau = 75$ slots, while similar DoF gains can be obtained using only $\tau=27$ slots, and also outperforming any other scheme. This is one of the main conclusions obtained from our analysis: while the best DoF are attained using a high number of time slots, usually one solution with reduced number of time slots can be found without high DoF penalties.
The reader may have noticed that no case with $\rho > K-1$ has been considered because the scheme in $\cite{Hao_BSR_MISOIC}$ surpasses the proposed TG scheme. One example for $K=6$ will be addressed in \refbox{Section}{sec:order}.

\subsection{PSR scheme}
\label{sec:BSRsens}

The performance of the PSR scheme is compared to the RIA scheme for $K=3$ users. Since the region of most interest for this scheme is $\rho > \frac{4}{5}$, we consider two representative antenna settings:  $(M,N)=(4,5)$, and SISO ($M=N=1$). In this case, following the expressions given in \refbox{Section}{sec:PSR} it is easy to see that:
\begin{IEEEeqnarray}{c}
 S_1^* \! =  
 \left\lceil 
    \frac{\varphi_1 + b }{N}
 \right\rceil =
 \left\lceil 
    \frac{5}{4}\frac{b}{N}
 \right\rceil ,\nonumber\\[3mm]
 S_2^* = 
 \left\lceil 
    \frac{\varphi_1 + \varphi_2 }{N}
 \right\rceil =
 \left\lceil 
    \frac{b}{3N}
 \right\rceil =
 \left\lceil 
    \frac{2\varphi_2 + \varphi_3 }{N}
 \right\rceil
 =  S_3^* ,\nonumber\\[3mm]
  d_j(\Bmax) = \underset{ b \leq \Bmax}{\text{maximize}} \,\, \frac{1}{N} 
  \frac{b}{S_1^*+ 3S_2^* +S_3^*}. \label{eq:fBSR}
\end{IEEEeqnarray}
The performance for the two settings is depicted in \refbox{Fig.}{fig:DoF_Bmax}-bottom. The most remarkable result is that whenever $\Bmax$ is below $b^*$, the RIA outperforms the PSR scheme. Moreover, notice that for the unbounded case a similar DoF performance (from 0.387 to 0.375) is obtained for RIA w.r.t. the PSR scheme with only a quarter of the number of slots (from 31 to 8).

\subsection{DoF with limited order of symbols}
\label{sec:order}

In order to compare the DoF-delay trade-off of the proposed schemes for $K>3$ to other approaches in the literature, the DoF are depicted for different values of the maximum order of symbols $\Theta$ in \refbox{Fig.}{fig:DoF_order}. Two settings for $K=6$ are considered: SISO at left, and $(M,N)=(6,1)$ (MISO) at right. First, since the scale may be confusing, it is worth to remark that the first operation point of the PSR scheme outperforming the RIA scheme requires 1154 slots ($\Theta=3$), in contrast to the 160 slots required by the latter. Also, it is remarkable how the number of slots grow when the order of the transmitted symbols is not limited, with negligible DoF gains. For example, when $\Theta=4$ the achievable DoF require a quarter of the unbounded case communication delay (from 39258 to 7898), and provide a 95\% of the unbounded case achievable DoF. 

For the MISO case, the supremacy of the proposed schemes in terms of practical terms is evident. While the TG scheme requires only 21 slots, the first operation point outperforming its DoF performace ($\Theta=4$) requires 495 slots. Also, notice that the gains from this latter point w.r.t. the unbounded case are negligible, while the number of slots increase threefold.

As a conclusion, it is observed that in pursuit of approaching the DoF outer bound it is better to increase the number of phases and the order to the transmitted symbols. However, when practical issues come into play, it is preferable to penalize the achievable DoF for the sake of complexity and communication latency.
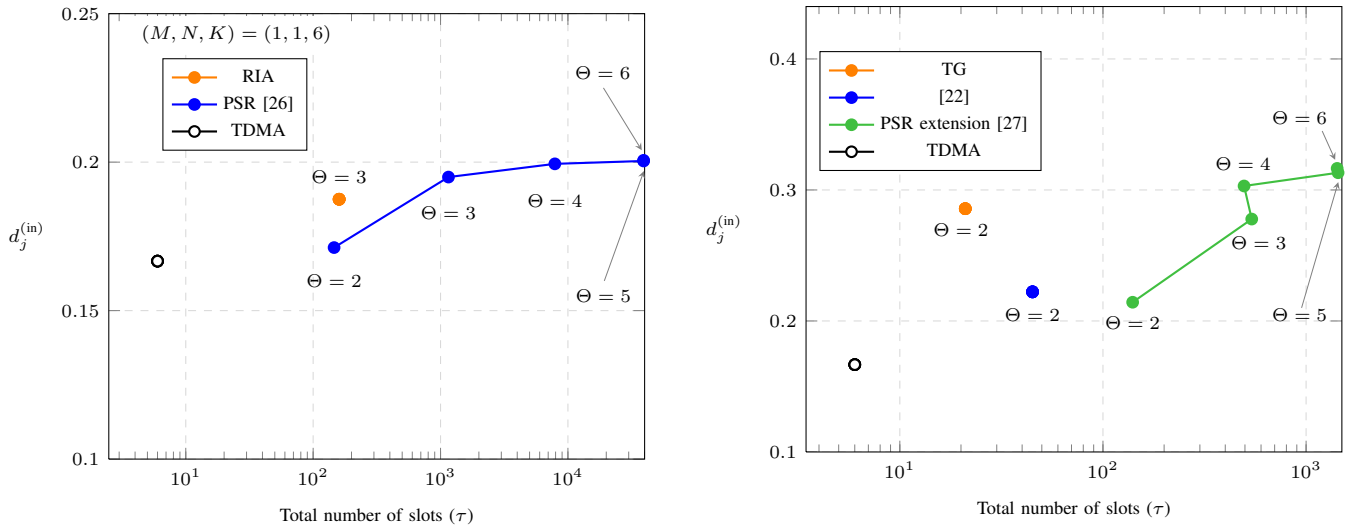
\begin{figure}
 
\begin{minipage}{\linewidth}
\begin{minipage}[]{0.48\linewidth}
  \centering 


  \newcommand{\dmin}{0.1}  \newcommand{\dmax}{0.25}
  \newcommand{\taumin}{0} \newcommand{\taumax}{40000}
  \newcommand{\conf}{116} 
\pgfplotstableread[col sep=tab]{DoF\conf.txt} \dof

\begin{tikzpicture} 
\begin{semilogxaxis}[
  ymin=\dmin,ymax=\dmax,xmin=\taumin,xmax=\taumax,
  xmajorgrids,
   ymajorgrids,
   grid style={dashed, gray!30},
  ylabel style={at={(0.02,0.5)},rotate=-90},
  width=\linewidth,   font=\scriptsize,
  xlabel=Total number of slots ($\tau$),  ylabel=$\djin$,
  legend style={at={(0.1,0.9)},anchor=north west,font=\scriptsize}],
  ]
  
   \addplot[plotoptsPropIn] 
  table[x =Rsum, y = d, restrict x to domain = 1:\taumax]   from \dof ; 
  \addlegendentryexpanded{RIA};  
    \addplot[plotoptsPrevIn] 
  table[x =Rsuma, y = da, restrict x to domain = 1:\taumax]   from \dof ; 
  \addlegendentryexpanded{PSR \cite{Abdoli_IC}};
  \addplot[plotoptsNoCSIT] 
  table[x =RsumnoCSIT, y = dnoCSIT]   from \dof ;   
  \addlegendentryexpanded{TDMA};
  
  \textboxx{160}{0.195}{$\Theta=3$} 
  \textboxx{146}{0.16}{$\Theta=2$} 
  \textboxx{1154}{0.183}{$\Theta=3$} 
  \textboxx{7898}{0.187}{$\Theta=4$} 
  \textboxx{18800}{0.23}{$\Theta=6$} 
  \linia{19000}{0.225}{38000}{0.204} 
  \textboxx{19000}{0.155}{$\Theta=5$} 
  \linia{19300}{0.16}{39500}{0.197} 
%

  \textbox{25}{0.243}{\scriptsize $(M,N,K)=(1,1,6)$}
  
\end{semilogxaxis}
\end{tikzpicture} 
\end{minipage}  \hspace{3mm}
\begin{minipage}{0.48\linewidth}
  \centering 


  \newcommand{\dmin}{0.1}  \newcommand{\dmax}{0.44}
  \newcommand{\taumin}{0} \newcommand{\taumax}{1500}
  \newcommand{\conf}{616} 
\pgfplotstableread[col sep=tab]{DoF\conf.txt} \dof

\begin{tikzpicture} 
\begin{semilogxaxis}[
  ymin=\dmin,ymax=\dmax,xmin=\taumin,xmax=\taumax,
  xmajorgrids,
   ymajorgrids,
   grid style={dashed, gray!30},
  ylabel style={at={(0.02,0.5)},rotate=-90},
  width=\linewidth,   font=\scriptsize,
  xlabel=Total number of slots ($\tau$),  
  ylabel=$\djin$,
  legend style={at={(0.025,0.9)},anchor=north west,font=\scriptsize}],
  ]
  
   \addplot[plotoptsPropIn] 
  table[x =Rsum, y = d, restrict x to domain = 1:\taumax]   from \dof ; 
  \addlegendentryexpanded{TG};
  \addplot[plotoptsPrevIn] 
  table[x =Rsumv, y = dv, restrict x to domain = 1:\taumax]   from \dof ; 
  \addlegendentryexpanded{\cite{Vaze2IC}};
  \addplot[plotoptsUp,green!50!gray, mark options={fill=green!50!gray,solid}] 
  table[x =Rsumh, y = dh, restrict x to domain = 1:\taumax]   from \dof ; 
  \addlegendentryexpanded{PSR extension \cite{Hao_BSR_MISOIC}};
  \addplot[plotoptsNoCSIT] 
  table[x =RsumnoCSIT, y = dnoCSIT]   from \dof ;   
  \addlegendentryexpanded{TDMA};
  
  \textboxx{45}{0.205}{$\Theta=2$} 
  \textboxx{20}{0.27}{$\Theta=2$}  
  \textboxx{140}{0.199}{$\Theta=2$} 
  \textboxx{585}{0.26}{$\Theta=3$}  
  \textboxx{490}{0.32}{$\Theta=4$}  
  \textboxx{930}{0.355}{$\Theta=6$} 
  \linia{1200}{0.345}{1390}{0.323}
  \textboxx{930}{0.205}{$\Theta=5$} 
  \linia{950}{0.21}{1440}{0.306} 
  
%
   
\end{semilogxaxis}
\end{tikzpicture} 
\end{minipage}  
\end{minipage}
\caption{Achievable DoF of the proposed schemes vs duration of the transmission $\tau$ for different values of the maximum order of the transmitted symbols $\Theta$. The DoF achieved without the need of CSIT or using previous schemes in the literature are also depicted for comparison purposes.}
\label{fig:DoF_order}
\end{figure}

\section{Achievable DoF for constant channels}
\label{sec:constant}

The literature on delayed CSIT always assumes that channel feedback incurs a delay larger than channel coherence time, i.e. the current channel is completely uncorrelated w.r.t. the channel that has been reported. However, this assumption is not always realistic in practice, since the transmitter has no way to know if the channel has changed. In this regard, this section studies the extreme case where the channel is constant, the transmitter is not aware of this, and performs a delayed CSIT strategy anyways. Then, the next sections prove \refbox{Theorem}{th:constant}, stating that all results so far also apply for constant channels.

The difference in the system model between constant and time-varying channels is that all block diagonal compositions of channels are simplified to Kronecker products. Let $\tilde{\Hn}_{j,i} \Cmat{N}{M}$ denote the channel between $\Tx{i}$ and $\Rx{j}$ for all $\tau$ slots of the communication, since the channels are constant. Then, we have
\begin{IEEEeqnarray*}{c}
 \Hn_{j,i} = \I_{\tau} \otimes \tilde{\Hn}_{j,i}.
\end{IEEEeqnarray*}
It is instructive to particularize it to the SISO case, where channels become scaled identity matrices, i.e:
\begin{IEEEeqnarray}{c}
 \Hn_{j,i} = \I_{\tau} \otimes \tilde{h}_{j,i} =\tilde{h}_{j,i} \I_{\tau},
 \label{eq:SISOch}
\end{IEEEeqnarray}
which exhibits lower diversity than MIMO channels. 


\subsection{RIA scheme}
\label{sec:RIAconst}

This section proves that the RIA scheme described in Section \ref{sec:innerboundRIA} fails for the SISO case if channels are constant and $L=3$. Next section will show that this scheme can be made feasible by means of exploiting asymmetric complex signaling concepts. Similar arguments allow showing feasibility for any other antenna setting with probability one. 

During the first phase of the RIA scheme, all transmitters are active, using predetermined precoding matrices $\Vn_i^{(1)}\Cmat{5}{3}$, and interfering to all users. The received signal is processed using the per-phase linear filters $\Un_{i,j} \Cmat{2}{5}$, in such a way that the desired signals are only mixed with interference from another user. Consider the signal space matrix for the signals received during the first phase:
\begin{IEEEeqnarray}{c}
 \Gn_i^{(1)} = \begin{bmatrix}
              \Un_{i,i+1} h_{i,i}\Vn_i^{(1)} & \Un_{i,i+1} h_{i,i+1}\Vn_{i+1}^{(1)} & \0 \\ \Un_{i,i-1} h_{i,i}\Vn_{i-1}^{(1)} & \0 & \Un_{i,i+1} h_{i,i-1}\Vn_{i-1}^{(1)}  \\
             \end{bmatrix}
             \label{eq:SSM_RIAconstant}
\end{IEEEeqnarray}
where indices in this section are assumed to be in the set $\{1,2,3\}$, applying the modulo-3 operation only if necessary. Notice that matrices $\Un_{i,j}$ satisfy
\begin{IEEEeqnarray}{c}
 \RSpan{\Un_{i,j}} = \Null\left({\Span{\Vn_k^{(1)}}}\right), \, \forall i,j \neq k, i \neq j, \label{eq:Ufilter}
\end{IEEEeqnarray}
i.e. $\Un_{i,j}$ removes the interference generated at $\Rx{i}$ by user $k\neq j$, but not the interference from user $j$. Due to definition (\ref{eq:Ufilter}), there are only three different per-phase filters. Indeed, they correspond to the null space of each ${\Vn}_{i}^{(1)}$, which will be denoted as $\bar{\Vn}_{i}^{(1)} \Cmat{2}{5}$ for ease of description. Accordingly, the signal space matrix for the whole communication writes as
\begin{IEEEeqnarray}{c}
 \Gn_i = \left[ \,\begin{matrix}
              h_{i,i} \bar{\Vn}_{i-1}^{(1)} \Vn_i^{(1)} & h_{i,i+1} \bar{\Vn}_{i-1}^{(1)} \Vn_{i+1}^{(1)} & \0 \\
              h_{i,i} \bar{\Vn}_{i+1}^{(1)} \Vn_{i-1}^{(1)} & \0 & h_{i,i-1} \bar{\Vn}_{i+1}^{(1)} \Vn_{i-1}^{(1)}  
              \\[1mm]  \cdashline{1-3}[3pt/2pt]  \\[-6.5mm] 
              h_{i,i} \Sigman_i^{(2)} \Tn_i^{(2)} &
              h_{i,i+1} \Sigman_{i+1}^{(2)} \Tn_{i+1}^{(2)} &
              h_{i,i-1} \Sigman_{i-1}^{(2)} \Tn_{i-1}^{(2)}
             \end{matrix} \, \right],
             \label{eq:SSM_RIAsimplified}
\end{IEEEeqnarray}
where the precoding matrices for the second phase are computed following (\ref{eq:scA:RIA_simple}) and (\ref{eq:scA:Tdef}), here repeated for reader's convenience:
\begin{IEEEeqnarray*}{c}
 \mathbf{V}_{i}^{(2)}  =  \boldsymbol{\Sigma}_i^{(2)} \Tn_{i}^{(2)} ,  \\[2mm]
 \RSpan{\Tn_i^{(2)}} = \Tc_i^{(2)} =  \Tc_{i+1,i} \cap \Tc_{i-1,i}  ,
\end{IEEEeqnarray*}
where $ \Tn_{j,i} \Cmat{2}{3}$, with $\Tn_{i+1,i}= h_{i+1,i} \bar{\Vn}_{i-1}^{\left(1\right)}  \Vn_{i}^{\left(1\right)} $ and $\Tn_{i-1,i}= h_{i-1,i} \bar{\Vn}_{i+1}^{\left(1\right)}  \Vn_{i}^{\left(1\right)} $. This design allows that the interference generated during the RIA phase be aligned with the JIS phase overheard interference at both non-intended receivers. 
Now, since $\Tn_{i+1,i}$ and $\Tn_{i-1,i}$ are independent, its intersection will be of dimension one with probability one. Then, there exist two vectors  $\un_i,\wn_i \Cmat{2}{1}$ such that $\Tn_i^{(2)}$ can be written as 
\begin{IEEEeqnarray}{c}
 \Tn_i^{(2)} \doteq 
 \un_i^T \bar{\Vn}_{i-1}^{(1)} \Vn_i^{(1)} \doteq
 \wn_i^T \bar{\Vn}_{i+1}^{(1)} \Vn_i^{(1)}, \label{eq:u1w1x}
\end{IEEEeqnarray}
%
where $\doteq$ is short for equality of row spans. Notice that $\un_i$ and $\wn_i$ correspond to the vectors that project $\bar{\Vn}_{i-1}^{(1)} \Vn_i^{(1)}$ and $\bar{\Vn}_{i+1}^{(1)} \Vn_i^{(1)}$ to its intersection subspace, respectively. The following lemma states a key property satisfied by these vectors:

\begin{minipage}{\linewidth}
\vskip 5mm
\begin{lemma}
If the vectors $\un_i,\wn_i$, $i=1,2,3$ are computed satisfying the properties in (\ref{eq:u1w1x}), then $\un_i \doteq \wn_{i+1}$. \\
 \label{lm:RIAconstant} 
\end{lemma}
\end{minipage}
\begin{IEEEproof}  
 Only the proof for $i=1$ will be shown. The proof for $i=2,3$ follows the same steps thus it is omitted.
 First, notice that (\ref{eq:u1w1x}) for $i=1,2$ can be written as follows:
 \begin{IEEEeqnarray}{c}
    \un_1^T \bar{\Vn}_{3}^{(1)} - \wn_1^T \bar{\Vn}_{2}^{(1)}
 \subset \bar{\Vn}_1^{(1)} \quad \Rightarrow \quad 
  \un_1^T \bar{\Vn}_{3}^{(1)} - \wn_1^T \bar{\Vn}_{2}^{(1)} = 
    \boldsymbol{\lambda}^T \bar{\Vn}_{1}^{(1)}, \\[2mm]
 \wn_2^T \bar{\Vn}_{3}^{(1)} - \un_2^T \bar{\Vn}_{1}^{(1)}
 \subset \bar{\Vn}_2^{(1)} \quad \Rightarrow \quad
 \wn_2^T \bar{\Vn}_{3}^{(1)} - \un_2^T \bar{\Vn}_{1}^{(1)} =
 \boldsymbol{\varphi}^T \bar{\Vn}_2^{(1)} .
 \end{IEEEeqnarray}
for some $\boldsymbol{\lambda}, \boldsymbol{\varphi} \Cmat{2}{1}$, which is equivalent to
 \begin{IEEEeqnarray}{c}
   \begin{bmatrix}
   \boldsymbol{\lambda}^T, & \wn_1^T , & \un_1^T
   \end{bmatrix}
   \begin{bmatrix}
    \bar{\Vn}_{1}^{(1)} \\
    \bar{\Vn}_{2}^{(1)} \\
    -\bar{\Vn}_{3}^{(1)}
   \end{bmatrix} = \0,
  \quad 
 \begin{bmatrix}
   \un_2^T,  & \boldsymbol{\varphi}^T, & \wn_2^T
   \end{bmatrix}
   \begin{bmatrix}
    \bar{\Vn}_{1}^{(1)} \\
    \bar{\Vn}_{2}^{(1)} \\
    -\bar{\Vn}_{3}^{(1)}
   \end{bmatrix} = \0.
 \end{IEEEeqnarray}
Hence, $\un_1$ and $\wn_2$ are the last two components of any vector lying on the null space of the $6 \times 5$ full rank matrix on the right hand side. Since it has dimension one, the last two components will always be proportional, thus $\un_1 \doteq \wn_2$. \\
\end{IEEEproof}   
\vskip 3mm
%

Linear feasibility requires that the rank of the equivalent channel is equal to the number of transmitted symbols. This will be settled in the negative for user one, while non-feasibility for the rest of users may be similarly proved. In this regard, consider its equivalent channel:
\begin{IEEEeqnarray}{c}
 \Hn_1^{(\text{eq})}  \sobreq{\text{(a)}}{=} \Wn_1 \Un_1 \Hn_1 \Vn_1  \sobreq{\text{(b)}}{=} \Wn_1 
 \left[ \, 
 \begin{matrix}
        h_{1,1} \bar{\Vn}_{3}^{(1)} \Vn_1^{(1)}    \\[1mm]
        h_{1,1} \bar{\Vn}_{2}^{(1)} \Vn_{1}^{(1)}  \\[1mm]
        \cdashline{1-3}[3pt/2pt]  \\[-6.5mm]
        h_{1,1} \Sigman_1^{(2)} \Tn_1^{(2)}
 \end{matrix} 
 \, \right],
 \label{eq:HeqRIAconst}
 \end{IEEEeqnarray} 
 where (a) is just a remainder of the definition (\ref{eq:HeqDef}) for the sake of reader's convenience, and (b) simply writes the equivalent channel as the first block column block rows of the signal space matrix (containing the desired signals) multiplied by the receiving filter $\Wn_1$. The objective of this filter is to remove the interference by combining the rows of the signal space matrix. One simple solution is
 \begin{IEEEeqnarray}{c}
  \Wn_1 = \begin{bmatrix}
           \Sigman_2^{(2)} \wn_2^T
           & 
           \Sigman_3^{(2)} \un_3^T 
           & 
           \I
          \end{bmatrix},
 \end{IEEEeqnarray}
thus the equivalent channel in (\ref{eq:HeqRIAconst}) writes as
\begin{IEEEeqnarray}{r l}
\Hn_1^{(\text{eq})}  & \doteq
       h_{1,1} \Sigman_2^{(2)} \wn_2^T \bar{\Vn}_{3}^{(1)} \Vn_1^{(1)} +
       h_{1,1} \Sigman_3^{(2)} \un_3^T \bar{\Vn}_{2}^{(1)} \Vn_1^{(1)} +
       h_{1,1} \Sigman_1^{(2)} \un_1^T \bar{\Vn}_{3}^{(1)} \Vn_1^{(1)}. \label{eq:Heq1}
\end{IEEEeqnarray} 
{First, note that $\un_1 \doteq \wn_{2}$ according to Lemma \ref{lm:RIAconstant}, thus the first and last terms are proportional. Moreover, note that the last term can be written as $h_{1,1} \Sigman_1^{(2)} \wn_1^T \bar{\Vn}_{2}^{(1)} \Vn_1^{(1)}$ due to definition (\ref{eq:u1w1x}). Then, since $\un_3 \doteq \wn_{1}$  holds according to Lemma \ref{lm:RIAconstant}, it is concluded that all three terms are proportional. Consequently, the equivalent channel has rank one, and the three desired symbols cannot be retrieved.}

\subsection{RIA scheme with ACS}
\label{sec:RIAconstACS}

As for the full CSIT case \cite{ACS}\cite{TAV_ppM1}, the application of asymmetric complex signaling concepts enables the feasibility of the RIA scheme either for constant or time-varying channels also for the SISO case. To the best of the authors knowledge, this is the first claim that improper signaling may be useful for precoding schemes using delayed CSIT. This section provides a sketch of the proof, omitted due to redundancy with the cited references.

In case of using asymmetric complex signaling, the channel can be modeled in terms of real magnitudes (see \cite{TAV_ppM1}), such that $2b$ {\it real symbols} are transmitted to each user along $2\tau$ slots, and the channel model in (\ref{eq:SISOch}) translates to
\begin{IEEEeqnarray}{c}
 \Hn_{j,i} = \I_{\tau} \otimes \big| \tilde{h}_{j,i} \big| \tilde{\Phin}_{j,i} 
 = \big| \tilde{h}_{j,i} \big| \Phin_{j,i} \Rmat{2\tau}{2\tau},
\end{IEEEeqnarray}
where $ \phi _{j,i}$ is the phase of the complex channel gain
$\tilde{h}_{j,i}$, and
\begin{IEEEeqnarray}{c}
\label{eq:matrixU}
\tilde{\Phin}_{j,i}  = \begin{bmatrix}
	\cos \left(  \phi _{j,i} \right) 
	& - \sin \left( {{\phi}_{j,i}} \right)
	\vspace{1mm}\\
	\sin \left(  {\phi} _{j,i} \right)
	& \cos \left( { \phi _{j,i}} \right)
\end{bmatrix} \Rmat{2}{2} , \\[3mm]
  \Phin_{j,i} = \I_{\tau} \otimes \tilde{\Phin}_{j,i}.
\end{IEEEeqnarray}
Matrices $\Phin_{j,i}$ break the diagonal structure of channel matrices. This is of interest because in previous section the same interference was generated at both unintenteded receivers thereby the same per-phase filter was used to remove it, see (\ref{eq:Ufilter}). Nonetheless, in this case different per-phase filters should be used, thus the connections among vectors $\un_i, \wn_i$ stated by \mbox{Lemma \ref{lm:RIAconstant}} no longer hold, and feasibility is ensured for any channel realization. Similar arguments apply to the MIMO case.

\subsection{TG scheme}
\label{sec:TGconst}

We review the foundations of this scheme, proposed in Section \ref{sec:innerboundTG} for $M>N$, in order to show that it also works for constant channels.
During the IIS phase transmitters are scheduled in a TDMA fashion. Therefore, for each $\Rx{j}$ obtains
\begin{IEEEeqnarray*}{c}
\yn_j^{\left(1,r\right)} =  \Hn_{j,r}^{\left(1,r\right)}  
 \Vns{r}{1,r}  \mathbf{x}_r = \left(\I_{S_1} \otimes \tilde{\Hn}_{j,i} \right)
 \Vns{r}{1,r}  \mathbf{x}_r, \\
 \Tn_{j,i} = \left(\I_{S_1} \otimes \tilde{\Hn}_{j,i} \right)  \Vns{i}{1,i},
\end{IEEEeqnarray*}
where the precoding matrices $\Vns{i}{1,i} \Cmat{MS_1}{b}$ are chosen to be some generic full-rank matrices, with $\Vns{i}{1,r}=\0$ for $r\neq i$. Since $M>N$, and $NS_1 < b$ by design, it is easy to see that all ranks are preserved even for constant channels, i.e.
$\Rank{\Tn_{j,i}}=NS_1, \, \forall i$, and all such pieces of overheard interference generate generic subspaces $\Tc_{j,i}$.

Now, let us recall that the precoders for each round of the RIA phase, see (\ref{eq:scB:Tdef}),  are linear combinations of $\Tn_i^{(2,r)}$, obtained as a basis of
\begin{IEEEeqnarray*}{c}
   \RSpan{\Tn_i^{(2,r)}} = \Tc_i^{(2,r)} =  \!\! \bigcap_{k\in \Ac^{(2,r)}\backslash \{i\} }  \!\!\!\!\Tc_{k,i}  
\end{IEEEeqnarray*}
which will also preserve the rank. Therefore, we conclude that this scheme does not require the time-varying channels assumption, since each receiver can acquire enough linear combinations of desired symbols even in case of constant channels.

\subsection{3-user PSR scheme}
\label{sec:BSRconst}
 
The first phase of this scheme is similar to that for the RIA scheme. In contrast, there are three phases and the second phase is developed by pairs. Feasibility is easily shown for MIMO channels, whereas the SISO setting fails. Since the scheme delivers exactly $12$ LCs of the $b=12$ desired symbols to each receiver, by simply showing that some of those LCs are linearly dependent is sufficient to show the non-feasibility. 
In this regard, next we show that not all LCs delivered during the first round of the second phase are linearly independent. Consider the signal space matrix for the second phase, particularized for this case:
\begin{IEEEeqnarray}{c}
\def\arraystretch{1.8}
\Gn_1^{(2)} = \left[ \,
\begin{matrix}
  h_{1,1} \bar{\Vn}_{3}^{(1)} \Vn_{1}^{(1)}  & \Tn_{1,2} &     \0          \\ 
   h_{1,1} \bar{\Vn}_{2}^{(1)} \Vn_{1}^{(1)} &      \0     & \Tn_{1,3}    
   \\[1mm]  \cdashline{1-3}[3pt/2pt]  \\[-6.5mm] 
  h_{1,1} \Sigman_1^{(2,1)} \Tn_{2,1} & 
  h_{1,2} \Sigman_2^{(2,1)} \Tn_{1,2} &  \0 \\
   h_{1,1} \Sigman_1^{(2,2)} \Tn_{3,1} & \0 &  
   h_{1,1} \Sigman_3^{(2,2)} \Tn_{1,3}\\
   \0  &  \Fng{3,2}{1} & \0 \\
   \0  &      \0       & \Fng{2,3}{1}\\
\end{matrix} \, \right],
\label{eq:BSR:matrixG2}
\end{IEEEeqnarray}
where the same notation as for the RIA case has been used, and in this case we have $\Tn_{2,1} = h_{2,1} \bar{\Vn}_{3}^{(1)} \Vn_{1}^{(1)} \Cmat{3}{12}$.

Two methods for delivering LCs of desired symbols were used in the second phase, see Section \ref{sec:PSR}. First, recall that $\Sigman_1^{(2,1)} \Cmat{4}{3}$, thus zero-forcing the interference received during the first round of the second phase, $\Rx{1}$ obtains
\begin{IEEEeqnarray}{c}
 \boldsymbol{\lambda}^T \, h_{1,1} \Sigman_1^{(2,1)} \Tn_{2,1} \xn_1
\end{IEEEeqnarray}
for some $\boldsymbol{\lambda} \Cmat{4}{1}$ that satisfies $\boldsymbol{\lambda}^T {\Sigman}_2^{(2,1)} = \0$. Clearly, such LC of desired signals lies on $\RSpan{\Tn_{2,1}}$.

On the other hand, four LCs of desired signals may be obtained by combining the JIS phase received signals with the signals received during the first round of the hybrid phase:
\begin{IEEEeqnarray}{c}
\left( h_{1,2} \Sigman_2^{(2,1)} h_{1,1} \bar{\Vn}_{3}^{(1)} 
\Vn_{1}^{(1)} + h_{1,1} \Sigman_1^{(2,1)} \Tn_{2,1} \right) \xn_1 = 
\left( \dfrac{ h_{1,2} } {h_{2,1}} \Sigman_2^{(2,1)}  + 
\Sigman_1^{(2,1)}  \right) h_{1,1} \Tn_{2,1} \xn_1.
\end{IEEEeqnarray}
Those LCs form a basis of the three-dimensional subspace $\RSpan{\Tn_{2,1}}$, thus actually would provide only three { \it independent} desired LCs of desired symbols. However, since the LC obtained by the first method lies also in $\RSpan{\Tn_{2,1}}$, after this round user one acquires only three instead of four {\it independent} desired LCs, and linear feasibility is discarded.

Nonetheless, this problem can be fixed by exploiting asymmetric complex signaling, since the per-phase receiving filters for the second phase are distinct across users, similarly to what occurs for the RIA scheme. Then, the PSR scheme can be made feasible even for SISO constant channels.

\section{Conclusion}
\label{sec:conclusion}
 
The DoF-delay trade-off has been studied for the $K$-user MIMO IC with delayed CSIT. Three fundamental tools are envisioned in the context of delayed CSIT for designing linear precoding strategies: {\it delayed CSIT precoding, user scheduling}, and {\it redundancy transmission}. Based on them, this work proposes three precoding strategies, evaluated as a function of the antenna ratio $\rho$.

For $\rho<1$, the RIA scheme initially proposed for the 3-user SISO IC ($\rho=1$) has been generalized to the $K$-user MIMO case. This scheme exploits {\it delayed CSIT precoding} and {\it redundancy transmission}. In contrast to the conjecture in \cite{RIA_Kusers}, our results show that state-of-the-art DoF can be improved by considering $L\geq 3$ active pairs. Moreover, we have shown that for the region $\frac{1}{K-1}<\rho<\frac{K}{K^2-K-1}$ our proposed inner bound using the RIA scheme gets very close to the best known outer bound.

Moreover, we have generalized the PSR scheme for 3 users from SISO to MIMO, which combines the three tools: {\it delayed CSIT precoding, user scheduling}, and {\it redundancy tranmission}. This scheme provides the best achievable DoF when the number of antennas at the transmitter and receiver are similar ($\rho \approx 1$) not only for the 3-user MIMO IC, but also for the $K$-user MIMO IC by applying time-sharing concepts. Nevertheless, a MIMO generalization for $K>3$ users remains open.

In case the transmitter has more antennas than the receiver ($\rho>1$), we propose the TG scheme improving state-of-the-art for $1 < \rho < K-1$. Linear precoding and user scheduling are carefully designed for DoF boosting, where the first phase is carried out orthogonally among users, whereas the second phase is developed in groups of $G\leq K$ users. The proper value of $G$ lies on the trade-off between the constraints imposed by interference alignment, and the increase on the number of rounds, in turn depending on the antenna ratio $\rho$ and the number of users $K$. 

The DoF-delay trade-off of the proposed schemes has been studied either by limiting the number or the order of the transmitted symbols. The first method builts upon the formulation of the parameters of each scheme (number of transmitted symbols and duration of the phases) as the solution of a DoF constrained maximization problem, and as a function of the number of users and the antenna ratio. In this regard, the analysis shows that although the PSR scheme and its extensions attain the best DoF values, this is at the cost of long transmission delays, which increases the complexity both at the transmitter and the receiver. 

Finally, the later part of this work has concluded that the time-varying channels assumption, which is common along all the literature on delayed CSIT, is indeed not necessary, except for the SISO case. This implies that delayed CSIT strategies can be used even if the channel remains constant, which could be the case if the transmitter does not actually know the current channel coefficients. For the particular SISO case, we have proved that the two schemes in the literature failed, which can be fixed by applying asymmetric complex signaling concepts. 

Many possible lines of future work remain open for this channel. On the one hand, a MIMO generalization of the PSR scheme (similar to the one in \cite{Hao_BSR_MISOIC}) for $K$ users may lead to tighter DoF results, although may be impractical in terms of communication delay.
On the other hand, {deriving tighter outer bounds would be desirable, since the achievable DoF of the best known schemes are still too far from the trivial upper bounds}. Finally, the formulation presented in this paper seems to be a good starting point for deriving precoding strategies for the assymmetric MIMO IC, i.e. when not all transmitters and receivers have the same number of antennas. In a similar way, it would be interesting to characterize not only the DoF per user or sum DoF, but also the DoF region for this channel.




\ifCLASSOPTIONcaptionsoff
  \newpage
\fi

\bibliographystyle{IEEEbib}
\bibliography{../../../papers/_referenciesMarc}

\end{document}